\documentclass[12pt,a4paper]{amsart}

\usepackage{empheq}
\usepackage{tcolorbox}
\usepackage{framed}
\usepackage{amsmath}
\usepackage[latin9]{inputenc}
\usepackage{amstext}
\usepackage{amssymb}
\usepackage{caption}
\usepackage{longtable}
\usepackage{lscape}
\usepackage{tabularx}
\usepackage{varioref}
\usepackage{amsthm}
\usepackage{graphicx}
\usepackage{txfonts}
\usepackage{pxfonts}
\usepackage{marginnote}
\usepackage{epic}
\usepackage{eepic}
\usepackage{float}
\usepackage{rotating}
\usepackage{epsfig}
\usepackage{indentfirst}
\usepackage{array}
\usepackage{varioref}
\usepackage{tikz}
\usepackage{marginnote}
\usepackage{pgf}
\usepackage{bbm}
\usepackage{appendix}
\usepackage{amsbsy}
\usepackage{latexsym}
\usepackage{amsfonts}
\usepackage{pstricks}
\usepackage{color}
\usepackage[numbers,square,comma, compress]{natbib}
\usepackage{eurosym}
\usepackage{tikz}
\usepackage{pdfpages}
\usepackage{wasysym}
\usepackage{caption}
\usepackage{subcaption}
\usepackage{braket}
\usepackage{simplewick}
\usepackage{yfonts}
\usetikzlibrary{backgrounds}

\usepackage[colorlinks=true,linkcolor=black,citecolor=black,urlcolor=black]{hyperref}

\usepackage[scale=0.85]{geometry}

\makeatletter

\pdfpageheight\paperheight
\pdfpagewidth\paperwidth

\numberwithin{equation}{section}
\numberwithin{figure}{section}

\numberwithin{equation}{section}
\numberwithin{figure}{section}

\let\emptyset\varnothing

\def\ll{\left\lgroup}
\def\rr{\right\rgroup}

\def\leq{\leqslant}
\def\geq{\geqslant}

\def\x{x^{(1)}}

\def\bref\textbf{\ref}

\def\ll{ \left\lgroup}
\def\rr{\right\rgroup}

\newcommand{\cC}{\mathcal{C}}

\newcommand{\cI}{\mathcal{I}}
\newcommand{\cJ}{\mathcal{J}}

\newcommand{\cL}{\mathcal{L}}
\newcommand{\cM}{\mathcal{M}}
\newcommand{\cN}{\mathcal{N}}
\newcommand{\cO}{\mathcal{O}}

\newcommand{\cW}{\mathcal{W}}

\renewcommand{\mod}{\textup{mod}\,}

\newcommand{\lt}{\left}
\newcommand{\rt}{\right}

\newcommand{\qbinom}[2]{{\genfrac{[}{]}{0pt}{}{#1}{#2}}_q}

\restylefloat{figure}

\usepackage[parfill]{parskip}

\begin{document} 

\begin{flushright}
    DIAS-STP-19-08
\end{flushright}
	
\title[]{
Closed form fermionic expressions for the Macdonald index}
\dedicatory{}
	
\author[]{Omar Foda      \!$^{{\scriptstyle {\, 1            }}}$ and 
          Rui-Dong Zhu   \!$^{{\scriptstyle {\, 2            }}}$
          }
\address{
$^{{\scriptstyle 1}}$
School of Mathematics and Statistics, 
The University of Melbourne, Parkville, Victoria 3010, Australia
\newline
$^{{\scriptstyle 2}}$
School of Theoretical Physics, 
Dublin Institute for Advanced Studies, 
10 Burlington Road, Dublin, Ireland
}

\email{
omar.foda@unimelb.edu.au,
nick\_zrd@stp.dias.ie
}

\keywords{
Argyres-Douglas theories,
Schur index,
Macdonalod index,
Non-unitary minimal models,
Fully-degenerate irreducible highest-weight modules,
Virasoro characters,
$W_N$ characters
}

\dedicatory{To Prof Barry M McCoy and \lq The Fermionic Characters\rq\\
In memory of Professor Omar Foda}
	
\begin{abstract}
We interpret aspects of the Schur indices, that were identified with characters 
of highest weight modules in Virasoro $(p,p')=(2,2k+3)$ minimal models for $k=1,2,\dots$, in terms 
of paths that first appeared in exact solutions in statistical mechanics. 
From that, we propose closed-form fermionic sum expressions, that is, 
$q, t$-series with manifestly non-negative coefficients, for  
two infinite-series of Macdonald indices of $(A_1,A_{2k})$ Argyres-Douglas theories 
that correspond to $t$-refinements of Virasoro $(p,p')=(2,2k+3)$ minimal model characters, 
and two rank-2 Macdonald indices that correspond to 
$t$-refinements of $\cW_3$ non-unitary minimal model characters. 
Our proposals match with computations from 4d $\mathcal{N} = 2$ gauge theories 
\textit{via} the TQFT picture, based on the work of J Song \cite{song.01}.\\

{\it My great collaborator, Prof. Omar Foda, passed away shortly after the completion of the third version of this paper. In fact, he shared the core idea of this paper with me back in 2017. It took me rather a long period to prepare and develop the tools we needed in this article, but I still feel lucky enough to finish this work with Omar. He first mentioned his illness to me in last September, when we started to summarize our results into an article. However, he was never willing to tell me too much about his health condition, and after certain times of surgeries, I thought he completely recovered from his illness, as he was so positive during our discussion on this paper. We even discussed about a lot of potential future directions after this work in December, and Omar also played the major role in the discussion with other research groups and the revision of this paper after we put it on the arXiv. I never thought he would leave us so soon.

Prof. Omar Foda was one of the most important people for me during my early academic life. He was also a kind and active collaborator for me. His brilliant ideas will continue to guide us in the future. May he rest in peace.} 

\begin{flushright}
{\it Rui-Dong Zhu}
\end{flushright}
\end{abstract}
	
\maketitle

\section{Introduction}
\label{section.01}

\subsection{Schur and Macdonald indices in Argyres-Douglas theories 
as vacuum and \texorpdfstring{$t$}{t}-refined vacuum \texorpdfstring{$W_N$}{wn} characters}
In \cite{beem.01, beem.02,cordova.shao}, Beem \textit{et al.} showed 
that the Schur indices in certain Argyres-Douglas theories are 
characters of irreducible highest-weight vacuum modules in a class of non-unitary 
$W_N$ minimal models.
In \cite{song.01}, Song proposed a method to compute the Macdonald 
indices that generalizes the Schur indices of \cite{beem.01, beem.02} 
as $q, t$-series expansions of $t$-refined 
irreducible highest-weight vacuum modules in the non-unitary Virasoro 
minimal models $\cM^{\, 2, \, 2 k + 3}$, $k = 1, 2, \cdots$. 

\subsection{Schur indices in Argyres-Douglas theories 
in the presence of surface operators}
In \cite{nishinaka.sasa.zhu}, Nishinaka \textit{et al.} studied 
the Schur indices 
in Argyres-Douglas theories in the presence of surface operators.
They considered two infinite series of Argyres-Douglas theories, 
\textbf{1.} the series labeled 
$\ll A_{n-1}, A_{m-1} \rr$ with ${\rm gcd}(n,m)=1$, and 
\textbf{2.} the series labeled 
$\ll A_{n-1}, A_{\, 2m} \rr$, for $n=2$, $m = 1, 2, \cdots$,
in the presence of the surface operator labeled by $s_i$, $i=1, \cdots, n-1$.
They showed that in these two infinite series, 
the Schur index matches the character of the $\cW$-algebra highest weight module 
with the same label $s_i$, $i=1, \cdots, n-1$. 
This generalizes the work of 
\cite{cordova.shao, song.01,buican.nishinaka.01,buican.nishinaka.02,buican.nishinaka.03,creutzig.01,creutzig.02}
on the vacuum modules, and the work of
\cite{cordova.gaiotto.shao1,cordova.gaiotto.shao2,cordova.gaiotto.shao3} 
on the non-vacuum modules, which also involves surface operators in gauge 
theory
\footnote{See \cite{song.xie.yan,bonetti.meneghelli.rastelli,xie.yan.01,pan.peelars.01,beem.maneghelli.rastelli,dedushenko.fluder,xie.yan.02,fluder.longhi,beem.maneghelli.peelaers.rastelli,xie.yan.03,dedushenko,pan.peelars.02,dedushenko.wang} 
for recent progress.
}.
In the present work, we focus on the first series whose dual is 
the $W_N$ minimal model labeled by $(p = n, p' = n + m)$
\footnote{\, 
In minimal models, usually the modules are labeled by positive integers $r_i$ 
and $s_i$, $i=1,\cdots,n$. However, due to the constraints $\sum_{i=1}^n r_i=p$ 
and $\sum_{i=1}^n s_i=p'$, for $p=n$, only $s_i$ for $i=1, \cdots, n-1$ are free 
parameters left. 
}. 
 
\subsection{Macdonald indices in Argyres-Douglas theories 
in the presence of surface operators}

In \cite{watanabe.zhu}, Watanabe \textit{et al.} extended the results 
of \cite{nishinaka.sasa.zhu} to the corresponding Macdonald indices. 
Sum expressions for the Macdonald indices were obtained in terms of 
Macdonald polynomials for the series 
$\ll A_{n-1}, A_{m-1} \rr$, ${\rm gcd}(n,m)=1$, 
for $n = 2, 3$, as a generalization of the results of \cite{song.02}. 
For $n = 2$, Macdonald indices  could be computed to arbitrary high 
orders, but for $n=3$, the Macdonald index was determined from this approach only to a high order 
$\ll \cO \ll q^{10} \rr \rr$. 
Due to the technical complication in the Higgsing method used in \cite{watanabe.zhu} 
to generate surface operators in Argyres-Douglas theories, only two infinite series 
of rank-2 Macdonald indices, 
the series that corresponds to the vacuum modules, and 
the series that corresponds to the next-to-vacuum modules of $\cW_3$ characters, 
were conjectured. 

\subsection{Virasoro characters as generating functions of weighted paths}
The local height probabilities in restricted solid-on-solid models 
(which are off-critical 1-point functions on the plane with specific boundary conditions) 
are generating functions of weighted paths
\cite{andrews.baxter.forrester, forrester.baxter}.
They are also equal to the characters of Virasoro minimal models
(which are critical partition functions on the cylinder with specific boundary conditions) 
\footnote{\,
The literature on this equivalence is extensive. For a comprehensive overview, discussion
and motivation, we refer the reader to \cite{foda.02}.
}, 
hence the latter have the same combinatorial interpretation as weighted paths.
There is more than one way to represent these weighted paths, and in this work, 
we adopt the representation of these weighted paths proposed in \cite{foda.welsh}.

The generating functions of these weighted paths admit more than one $q$-series 
representation. One of these representations is a constant-sign sum with manifestly 
non-negative coefficients. The coefficient $a_n$ of $q^{\, n}$ in this representation
is the multiplicity of the states of conformal dimension $n$ (up a possible 
shift common to all states) in the corresponding irreducible highest-weight module. 
In \cite{berkovich,berkovich.mccoy.01,berkovich.mccoy.schilling,foda.warnaar, 
kedem.klassen.mccoy.melzer, warnaar.01, warnaar.02, warnaar.03}
 these states were interpreted 
in terms of (quasi-)particles and their weights (the corresponding power of $q$) 
were interpreted in terms of their (quasi-)momenta. These manifestly non-negative
sum expressions were called \textit{\lq fermionic characters\rq}
\footnote{\, 
The papers \cite{warnaar.01, warnaar.02} focus on the unitary minimal models, 
using the combinatorics of the paths that are appropriate to the unitary models,
while \cite{foda.warnaar} completes the proof in this case.}
\footnote{\,
There is another approach to the fermionic characters using path algebras of 
fusion graphs 
\cite{kellendonk.recknagel,kellendonk.rosgen.varnhagen,kaufmann}
}

\subsection{Closed form expressions for the Macdonald index}
In \cite{foda.01}, and independently \cite{rastelli.private.communication}, it was noted that Song's 
$q, t$-series for the vacuum modules of $\cM^{\, 2, \, 2 k + 3}$, $k=1, 2, \cdots$ 
are generated by a specific $t$-refinement of the fermionic form of the corresponding 
Virasoro characters. In the present work, we extend and check this observation.

We show that 
\textbf{1.} aspects of the $\cW_2$ Schur index can be read directly from the 
paths, including 
the multiplicities of the Schur operators that contribute to the index, 
the composition of these operators in terms of Schur operators that are 
not derivatives of simpler ones 
(we call these \lq primary Schur operators\rq as defined in \ref{primary.schur}) 
and Schur operators that are derivatives of simpler ones 
(we call these \lq descendant Schur operators\rq as defined in \ref{primary.schur}), 
as well as the precise counting of the derivatives, 
and 
\textbf{2.} that a refinement of these sum expressions in terms of a parameter $t$ 
with a specific power that depends on the numbers of particles,
gives a closed form expression for the corresponding Macdonald character.
We match our results with direct computations from the Argyres-Douglas theory side,
based on a method proposed by J Song \cite{song.01} and find complete agreement
in cases where results are available from both sides.

\subsection{Outline of contents and results}
In section \textbf{\ref{sec:def-gauge}} and \textbf{\ref{sec:def-sta}}, 
we introduce basic definitions that we need in the sequel, from the gauge theory side 
and from the statistical mechanics side, respectively, including the superconformal 
index, the Schur operators, the fermionic forms of the characters of the Virasoro $(p,p')=(2,2k+3)$
non-unitary minimal models ($k=1,2,\dots$), as well a specific $\cW_3$ non-unitary minimal model. 
Based on the fermionic form of the characters, we review the quasi-particle picture 
of the Virasoro minimal models, and define natural $t$-refined characters for 
these models by assigning different $t$-weights to different particle species.
In section \textbf{\ref{sec:proposal}}, we conjecture that the $t$-refined character 
is equal to the Macdonald index computed from the gauge theory side, 
based on the observation that they match as series expansions in $q$, up to a high order. 
Next, we make the stronger conjecture that the quasi-particles of statistical 
mechanics are in one-to-one correspondence with the Schur operators that are 
counted by the Schur/Macdonald index in the gauge theory. 
Section \textbf{\ref{sec:comment}} contains a number of comments. 

\subsubsection{Remark}
\textit{
We focus on the Virasoro characters, two infinite
series of which are considered in this work. Following that, we discuss 
the case of two $\cW_3$ characters separately and in analogous terms.
}
\subsubsection{Remark}
\textit{
While we normally use the terminology $t$-refinements to add a parameter $t$, 
it is often convenient to think in terms of $T$-refinements instead where $T := t/q$. 
}

\section{Definitions. The gauge theory side}\label{sec:def-gauge}
\textit{We recall basic definitions from the gauge theory side}

\subsection{The 3-parameter superconformal index of 
4d \texorpdfstring{$\cN \, = \, 2$} {n=2} superconformal field 
theories}

The superconformal index is defined \cite{aharony.01,Kinney:2005ej} as the 3-parameter 
Witten index  

\begin{equation}
\cI(p,q,t) = 
{\rm tr}
\ll 
(-1)^F p^{\frac{E-2j_1-2R-r}{2}} q^{\frac{E+2j_1-2R-r}{2}} t^{R+r} e^{-\beta H}
\rr,
\label{index-pqt}
\end{equation}

where $(E,j_1,j_2,R,r)$ are the quantum numbers associated to the $\cN \, = \, 2$ 
superconformal algebra, that is, the dilatation charge, the spins, SU(2)$_R$ 
charge and U(1)$_r$ charge, $F$ is the fermion number and the Hamiltonian $H$ 
can be chosen as
\footnote{\, 
A review of the 4d $\cN \, =2 \, $ superconformal algebra can be for example found 
in \cite{beem.01}. $\{\bullet ,\bullet\}$ denotes the anti-commutator of 
fermionic operators. 
} 

\begin{equation}
H = 2 
\lt\{
\bar{\mathcal{Q}}_{1\dot{-}},\bar{\mathcal{Q}}^\dagger_{1\dot{-}}
\rt\}
= 2 \ll E-2j_2-2R+r \rr 
\end{equation}

The local operators that contribute to the superconformal index are BPS 
operators annihilated by $H$, or equivalently by $\bar{\mathcal{Q}}_{1\dot{-}}$. 

\subsection{The Schur operators of 4d \texorpdfstring{$\cN \, =2 \, $}{n=2} 
superconformal field theories} 

The superconformal index depends on three fugacity parameters, $p$, $q$ and $t$. One can 
consider some special limit of the index, where the Hilbert subspace contributing to the 
index is further restricted. The Macdonald limit, $p\rightarrow 0$, restricts the index 
to local operators that are not only annihilated by the Hamiltonian, but also satisfy 

\begin{equation}
\label{condition.01}
E-2j_1-2R-r=0, 
\end{equation}

or equivalently

\begin{equation}
E=j_1+j_2+2R,\quad r+j_1-j_2=0
\end{equation}

These are called Schur operators. We refer the readers to \cite{beem.01} for 
the conventions and discussions used here, with a (limited) list of possible 
Schur operators. 

\subsection{The chiral algebra of 4d \texorpdfstring{$\cN \, =2 \, $}{n=2} 
superconformal field theories}
In \cite{beem.01}, a systematic method was discovered to construct a chiral 
algebra spanned by the Schur operators of 4d $\cN \, =2 \, $ superconformal 
field theories. 
The dual chiral algebra contains the Virasoro algebra with central charge 
$c_{\, 2d}$ given by the $c$-coefficient, $c_{4d}$, in the 4-point function 
of stress tensors in 4d, as 

\begin{equation}
c_{\, 2d} = - 12 c_{\, 4d}
\end{equation}

\subsection{The Schur index}

The Schur index is the Schur limit, $p\rightarrow 0$, $q=t$, of the superconformal 
index and coincides with the character of the vacuum irreducible highest weight 
module of the corresponding chiral algebra

\begin{equation}
\cI(q)={\rm tr} \ll (-1)^F q^{\, h} \rr,
\end{equation}

where the conformal weight of a 2d chiral algebra state is 

\begin{equation}
h=R+j_1+j_2 , 
\end{equation}

in 4d terms.

\subsection{Primary and descendant Schur and \texorpdfstring{$W_N$}{Wn} operators}
\label{primary.schur}

In this work, we study isomorphisms between the Schur sector in 4d superconformal 
field theories and irreducible highest weight modules in 2d chiral algebras. 

Elements in the Schur sector (the set of all Schur operators) 
can be classified (as we show in the sequel) into a set of
finitely-many Schur operators that are not descendants of 
other Schur operators under the action of the 4d superconformal algebra
\footnote{\,
The property that there are finitely-many such operators may 
be true only in the Argyres-Douglas theories/minimal models 
studied in this work. It is possible that the Schur sectors 
of more general models have infinitely many primary Schur 
operators.
}, 
and a set of Schur operators that are descendants of other 
Schur operators under the action of the 4d superconformal 
algebra (that is, the action with derivatives on the first 
set of Schur operators).
In the sequel, we call the first type 
\textit{primary Schur operators}, and the second type
\textit{descendant Schur operators}
\footnote{\,
Note that this terminology is new, we introduce it for 
the purposes of this work.
}. 

An irreducible highest weight module in 
a 2d $W_N$ minimal conformal field theory consists 
(as well known) of  
a single highest weight state created by the action of a primary 
$W_n$ operator on the vacuum state (in the case of the vacuum 
highest weight module, the primary $W_n$ operator is the identity), 
and infinitely many descendant states that are generated by 
the action of $W_N$ operators on the highest weight state.
We call the first type
\textit{primary $W_N$ operators}, and the second type
\textit{descendant $W_N$ operators}
\footnote{\,
These are the known primary and descendant $W_N$ generators. 
We use $W_n$ when necessary to avoid confusion.
}.

Each Schur operator in a Schur sector of the type studied in 
this work 
is in bijection with a $W_N$ operator. However, since there 
are (as we will show) in general finitely-many primary Schur
operators in a Schur 
sector and one primary $W_N$ operator in 
a $W_N$ irreducible highest weight module, only one primary 
Schur operator maps to that primary $W_n$ operator, 
while the remaining primary Schur operators map to $W_n$ 
descendant operators. In the sequel, it is convenient to 
restrict the definition 
of primary Schur operators to those that map to descendant $W_N$ 
operators (in other words, we exclude the Schur operator that 
maps to the primary $W_N$ operator). 

As we show in the sequel, one of the results of this work is that  
the primary Schur operators (that map to descendant $W_N$ operators)  
are distinguished in the sense that they create the particles that 
make the spectrum of the 2d $W_N$ minimal conformal field theory. 

A known example of a primary Schur operator which maps to 
a $W_n$ descendant operator is the R-symmetry current in the 
stress tensor multiplet, which maps to a $W_n$ descendant operator 
of conformal weight $2$, under the state-operator correspondence
\footnote{\,
A primary Schur operator is not necessarily primary under the action 
of the full 4d $\cN=2$ superconformal algebra. For example, the R-symmetry 
current is a descendant of a scalar field.}. 

\subsection{The Macdonald index}

The Macdonald index is the Macdonald limit, $p\rightarrow 0$, of the superconformal 
index. As the same set of operators, the Schur operators, contribute to the Macdonald 
index, it is also related to the chiral algebra, as a one-parameter $t$-refined version 
of the character. 
In \cite{song.02}, Song found that the quantum number $\ell=R+r$ in the Macdonald index 

\begin{equation}
\cI(q,t) = {\rm tr} \ll (-1)^F T^{\, \ell} q^{\, h} \rr,
\label{ref-character}
\end{equation}

where 

\begin{equation}
T := t/q 
\end{equation}

counts the number of fundamental generators in the chiral algebra used to 
obtain each state starting from the highest weight. A more detailed review on Song's work will be provided in section \ref{s:Song}. 

\subsection{Argyres-Douglas superconformal field theories} In the case of a weakly-coupled 
superconformal gauge theory with a Lagrangian description, one can write a matrix integral 
based on the field content of the gauge theory, and using that, evaluate the superconformal 
index \cite{aharony.01}. 
An Argyres-Douglas theory is strongly-coupled and has no Lagrangian description. 
However, one can compute the superconformal index using the class S theory construction, 
that is the compactification of 6d $\cN= \ll 2, 0 \rr$ theory on a Riemann surface 
with an irregular puncture, and compute the index using the TQFT defined on the Riemann 
surface \cite{gadde:2011-1,song.01}. 
Further, in the case of rank-one Argyres-Douglas theories, it is not difficult to compute 
the index from BPS quivers \cite{cordova.shao} and the RG flow from 4d $\cN \, =2 \, $ SYM 
\cite{maruyoshi.song,maruyoshi.song.02,agarwal:2016}. 
In this work, we focus on Argyres-Douglas theories of type $\ll A_{n-1}, A_{m-1} \rr$,
${\rm gcd}(n,m)=1$.

\subsection{TQFT approach to Macdonald index}

The Macdonald index of the class of theories we study in this article can be computed 
\textit{via} the so-called TQFT approach as 

\begin{equation}
\cI_{\ll A_{n-1}, A_{m-1} \rr} (q,t) =
\sum_\lambda C^{-1}_\lambda (q,t) \, f^{I_{n,m}}_\lambda (q,t) ,
\label{Mac-from-TQFT}
\end{equation}

where $\lambda = \{\lambda_i\}_{i=1}^{n-1}$ is a partition with $n-1$ rows, $C_\lambda$ 
is the 3-pt coefficient in the TQFT picture

\begin{equation}
C^{-1}_\lambda(q,t)
=
\frac{\tilde{P}_\lambda(t^\rho;q,t)}{\prod_{i=1}^r (t^{d_i};q)_\infty},
\end{equation}

with $(a;q)_\infty=\prod_{i=0}^\infty(1-aq^i)$, $d_i$ is the degree of $i$-th Casimir in 
the Lie algebra $A_{n-1}$, $\tilde{P}_\lambda(x;q,t)$ is the normalized Macdonald polynomial 
of $A_{n-1}$-type that satisfies 

\begin{equation}
\frac{1}{n!} \frac{(q;q)^{n-1}}{(t;q)^{n-1}} \oint \prod_i 
\frac{{\rm d}z_i}{2\pi iz_i}
\prod_{\alpha\in\Delta}
\frac{(z^\alpha;q)}{(tz^\alpha;q)} \tilde{P}_\lambda(z;q,t) \, \tilde{P}_\mu(z^{-1};q,t) = \delta_{\lambda\mu},
\end{equation}

and $f^{I_{n,m}}_\lambda$ is the wavefunction of the irregular puncture $I_{n,m}$ 
\cite{song.01, watanabe.zhu}. For example, the wavefunction of $I_{2,2i+1}$ 
$i = 1, 2, 3, \cdots$, is 

\begin{equation}
f^{I_{2,2i+1}}_\lambda(q,t) = (-1)^{\frac{\lambda}{2}} q^{\frac{
\lambda}{2}(\frac{\lambda}{2}+1)(i+\frac{3}{2})}(t/q)^{\frac{\lambda}{2}(i+2)}
\ll  \frac{(t;q)_{\frac{\lambda}{2}}(q^{\frac{\lambda}{2}+1};q)_{\frac{\lambda}{2}}
(t^2q^{\lambda};q)
}{
(q;q)_{ \frac{\lambda}{2}} (tq^{\frac{\lambda}{2}};q)_{\frac{\lambda}{2}} (tq^{\lambda+1};q)} \rr ^{\frac{1}{2}} ,
\label{wave-rank-one}    
\end{equation}

where $\lambda$, a one-row partition, is even, and the wavefunction is zero when $\lambda$ is odd. 
For $I_{3,m}$, similarly, the wavefunction does not vanish only when the corresponding 
weight $\vec{w}$ of the representation $\lambda$ of $A_2$, that is $w_1=\lambda_1-\lambda_2$, $w_2=\lambda_2$, takes the form 
\begin{equation}
    (w_1,w_2)=(3k,3\ell),\quad{\rm or}\quad  (w_1,w_2)=(3k-2,3\ell-2),
\end{equation}
for some appropriate integers $k$ and $\ell$. For more details, refer to \cite{song.01,watanabe.zhu}

\subsubsection{Remark} 
\textit{
When we expand the index with respect to $q$, the contribution from each 
$f^{I_{n,m}}_\lambda$ to the index starts from the level $(n+m)h(\lambda,n)$, where $h(\lambda,n)$ 
is a function that is independent of $m$. 
For example, $h(\lambda,2)=\frac{\lambda}{4}(\frac{\lambda}{2}+1)$. 
In other words, if we truncate the index at for example $q^{10}$, the index for smaller $m$ 
contains more non-trivial information from the viewpoint of TQFT. 
}

\subsubsection{Remark} 
\textit{
Following \cite{watanabe.zhu}, the Macdonald index was shown to match exactly with 
the $t$-refined character of 
the vacuum and next-to-vacuum module in the large $m$ limit of $(p,p')=(n,n+m)$ minimal models. This also motivates us to focus on the case of small $m$ in this work. 
}
\subsection{Song's work}
\label{s:Song}

In \cite{song.02}, Song showed that the Macdonald index of $ \ll A_1, A_{2k-2} \rr$ theory, 
which is dual to the $\cW_2$ non-unitary $(p,p')=(2,2k+3)$ minimal model $\cM^{\, 2, 2k+3}$ \cite{cordova.shao}, 
is a $t$-refined character of Virasoro algebra that can be computed as follows. We first introduce 
the parameter $T:=t/q$. To each state in the module that can be written as 

\begin{equation}
L_{-i_1} L_{-i_2}
  \cdots L_{-i_m}
\ket{\, 0 \,},\quad {\rm with}\ \  i_1+i_2+  \cdots+i_m=h,
\end{equation}

we assign a weight $T^m q^{\, h}$, that is $\ell=m$ in (\ref{ref-character}), 
and the $t$-refined character is given by the sum of the contributions of all the states in 
the vacuum module of the dual chiral algebra. When there are null states in the module, we 
delete the states with largest $T$-weight from the spectrum. 

To access non-vacuum modules from the gauge theory side, one needs to either insert defect 
operators in the perpendicular direction to the chiral algebra plane in 4d \cite{cordova.gaiotto.shao1,cordova.gaiotto.shao2,cordova.gaiotto.shao3}, or consider the 
lens space index of the gauge theory \cite{fluder.song}. In the case of $\ll A_{n-1}, A_{m-1} \rr$ 
theories with ${\rm gcd}(n,m)=1$, the former approach is more powerful, and the correspondence 
between surface operators and non-vacuum modules of chiral algebra was worked out in 
\cite{nishinaka.sasa.zhu}. 

The Macdonald indices in higher-rank cases and with surface operator inserted in 
$\ll A_{n-1}, A_{m-1} \rr$ theories with ${\rm gcd}(n,m)=1$ are computed in 
\cite{watanabe.zhu}, \textit{via} the TQFT approach and the Higgsing method, 
introduced in \cite{gaiotto.rastelli.razamat}, to generate surface operators 
in gauge theory. We will not describe the details of the Higgsing method, but 
essentially what it does to the Macdonald index (\ref{Mac-from-TQFT}), in correspondence 
with inserting a surface operator (labeled by $a$) in the gauge theory, is to insert 
a factor of Macdonald polynomial. For example, for $n=2$

\begin{equation}
\mathfrak{S}^a_\lambda(q,t) = 
q^{\frac{a}{2}} P_{(a)} \ll t^{\frac{1}{2}}q^{\frac{\lambda}{2}} \rr,
\end{equation}

is inserted in the sum expression over $\lambda$. 
The study of Macdonald indices computed in this way suggests that the $t$-refined character 
in higher-rank $W_N$-algebras that reproduces the Macdonald index can be obtained as follows. 
Given a state generated from the highest weight state by $m_j$ spin-$j$ currents, $W^{(j)}$, 
we assign

\begin{equation}
    T^{\, \ell} q^{\, h}, \, \,  \ell = \sum_{j = 2}^n \, m_j,
\end{equation}

to that state as its contribution, and we sum over all possible contributions to obtain the
$t$-refined character. 

In this work, we take a different approach, namely, we start from a statistical 
mechanics model described in the next section, define a natural $t$-refined character 
to it, and compare the result with the Macdonald index. 

\section{Definitions. The statistical mechanics/combinatorics side}\label{sec:def-sta}
\textit{We recall basic definitions from the statistical mechanics/combinatorics side}

\subsection{Alternating-sign (bosonic) sum expressions of the Virasoro characters}

For a minimal Virasoro model labelled by $p, p', r, s$, $p < p'$, $0 < r , p$, 
$0 < s < p'$, the character can be written in 
the alternating-sign (Feigin-Fuchs) form 

\begin{equation}
\label{bosonic-equation}
\chi^{\, p,  p'}_{r, s}=
{
\frac{1}{(q)_\infty}
}
\sum_{\lambda = - \infty}^\infty
(
q^{\, \lambda^2 p p' + \lambda (p'r-ps)
}
-
q^{\, (\lambda p + r) (\lambda p' + s)}
),
\end{equation}

with $(q)_{\infty}=\prod_{i=1}^\infty (1-q^{i})$,
and the {\em conformal dimension} $\Delta^{\, p,  p'}_{r, s}$ is 

\begin{equation}\label{conformal-equation}
\Delta^{\, p,  p'}_{r, s}=
\frac{(p' r - p s)^2 - (p' - p)^2}{4 p p'}
\end{equation}

Expression \eqref{bosonic-equation} for $\chi^{\, p,  p'}_{r, s}$
is related to the free-boson realization of the Virasoro 
algebra, and is known as a {\em bosonic} expression.
For later purposes, it will be useful to note that
$\chi^{p,p'}_{r,s}=\chi^{p,p'}_{p-r,p'-s}$ and that
$\chi^{p,p'}_{r,s}|_{q=0}=1$. 

\subsection{Constant-sign (fermionic) sum expressions of the Virasoro 
characters}

For $\cL^{\, 2, 2k+3}$, a constant-sign (fermionic) sum expression of 
the Virasoro characters is 

\begin{equation}
\label{AG-1}
\chi^{\, 2, 2 k + 3}_{r = 1,s = a}(q)=
\sum_{N_1 \ge \cdots \ge N_{k} \ge 0}
\frac{ 
q^{ N_{1}^2 + \cdots + N_{k}^2 + N_{a} + \cdots + N_{k} }
}{
(q)_{N_{1}-N_{2}} \cdots (q)_{N_{k-1}-N_{k}} (q)_{N_{k}}
},
\end{equation}

where $|q|<1$, $(q)_0=1$ and $(q)_n=\prod_{i=1}^{n}(1-q^i)$ for $n>0$.
Here $k\ge1$ and $1\le a\le k+1$. These are the expressions that we focus 
on in this work. 

\subsection{Product expressions of the Virasoro characters}

The above fermionic character expressions (\ref{AG-1}) satisfy the Andrews-Gordon 
identities \cite{andrews.01,gordon} 

\begin{equation}
\label{AG}
\sum_{N_1 \ge \cdots \ge N_{k} \ge 0}
\frac{ 
q^{ N_{1}^2 + \cdots + N_{k}^2 + N_{a} + \cdots + N_{k} }
}{
(q)_{N_{1}-N_{2}} \cdots (q)_{N_{k-1}-N_{k}} (q)_{N_{k}}
}
=
\prod_{\begin{subarray}{c}
  n=1
  \\
  n \not \equiv0, \pm i \, (\mod 2 k + 3) 
  \end{subarray}}^{\infty}
\frac{1}{1-q^n}
\end{equation}

The $k=1$ cases are the Rogers-Ramanujan identities
\cite{rogers,rogers-ramanujan}.

\subsection{The work of Bressoud}

In \cite{bressoud}, Bressoud interpreted the fermionic sum expression 
(\ref{AG-1}) as the character of Dyck paths with fixed initial and end points. 
This interpretation works only in the case of $\cL^{\, 2, 2k+3}$ models, $k=1, 2, \cdots$.
An equivalent intrpretation, also in terms of Dyck paths, developed in \cite{foda.welsh}, 
extends to all $(p, p')$ Virasoro minimal models. In this work, we use the paths of 
\cite{foda.welsh}, a review of which is in the next subsection.

\subsection{The paths of Virasoro minimal model characters. The vacuum modules}

One can express a Virasoro minimal model character as the generating function of weighted 
Dyck paths that connect two given points on a restricted-height semi-infinite lattice. 
More precisely, for a $(p,p')$ model, one prepares a lattice which is $p'-1$ bands in 
height, and $L+2$ bands in length, and considers Dyck paths that connect the points 
$\{(i,h_i)\}_{i=0}^{L+1}$, and satisfy

\begin{itemize}
    \item $h_0=a$, $h_L=b$, $h_{L+1}=c$ ($c=b\pm 1$),
    \item $h_{i+1}=h_i\pm 1$
\end{itemize}

The correspondence with the Virasoro minimal model characters is obtained by choosing 
the labels $(r, s)$ of the characters such that $s = a$, and $r$ is  

\begin{equation}
    r=\lfloor p c / p'\rfloor+\frac{b - c + 1}{2}
    \label{r-def}
\end{equation}

A ground-state band is defined as a band between $j$-th line and $(j+1)$-th line,
such that 

\begin{equation}
    \lfloor jp/p'\rfloor \neq \lfloor (j+1)p/p'\rfloor
\end{equation}

For example, in the $\cL^{\, 2, 5}$ Lee-Yang model, we have a $4\times (L+2)$-lattice 
(see Figure \ref{figure.path.01}) and the ground-state band lies between the 2nd and 
the 3rd lines in the lattice. 
Further, we need to assign a coordinate system $(x,y)$ with

\begin{equation}
    x_i=\frac{i-(h_i-a)}{2},\quad y_i=\frac{i+(h_i-a)}{2},
\end{equation}

to each point $(i,h_i)$. 

\subsubsection{The weight of a path} 
\begin{figure}
    \begin{tikzpicture}
    \draw [thin] (0.0,0,0)--(0.0,2.0);
    \draw [thin] (0.0,0.0)--(2.0,0.0);
    \draw [thin] (0.0,2.0)--(2.0,2.0);
    \draw [thin] (2.0,2.0)--(2.0,0.0);
    \draw [thin] (1.0,0,0)--(1.0,2.0);
    \draw [thin] (0.0,1.0)--(2.0,1.0);
    
    \foreach \x in {0.0,...,2.0}
{
\draw [fill=black!50] (\x,2) circle (0.04);
\draw [fill=black!50] (\x,1) circle (0.04);
\draw [fill=black!50] (\x,0) circle (0.04);
}

\begin{scope}[on background layer]
\fill[gray!50] 
(0.00, 2.00) -- (2.00, 2.00) -- (2.00, 1.00) -- (0.00, 1.00);
\end{scope}

    \draw [thick] (0,0)--(2,2);
    \node [below] at (1,0) {$(a)$};
    
    \draw [thin] (4.0,0,0)--(4.0,2.0);
    \draw [thin] (4.0,0.0)--(6.0,0.0);
    \draw [thin] (4.0,2.0)--(6.0,2.0);
    \draw [thin] (6.0,2.0)--(6.0,0.0);
    \draw [thin] (5.0,0,0)--(5.0,2.0);
    \draw [thin] (4.0,1.0)--(6.0,1.0);
    
    \foreach \x in {4.0,...,6.0}
{
\draw [fill=black!50] (\x,2) circle (0.04);
\draw [fill=black!50] (\x,1) circle (0.04);
\draw [fill=black!50] (\x,0) circle (0.04);
}

\begin{scope}[on background layer]
\fill[gray!15] 
(4.00, 1.00) -- (6.00, 1.00) -- (6.00, 2.00) -- (4.00, 2.00);
\end{scope}

    \draw [thick] (4,0)--(5,1)--(6,0);
    \node [below] at (5,0) {$(b)$};

    \end{tikzpicture}
    \caption{\textit{
    The two cases where we assign the weight $c_i=x_i$ to the middle point 
    located at position $i$. Scanning from left to right,
    (a) the path enters the ground-state (shaded) band 
    from a non-ground-state (white) band below at the middle point, 
    (b) the path reaches a peak at the middle point outside 
    the ground-state (shaded) band. 
    A lightly-shaded band (the upper band in Figure (b)) can be a ground-state 
    band or not.
    }}
    \label{fig:weight-1}
\end{figure}
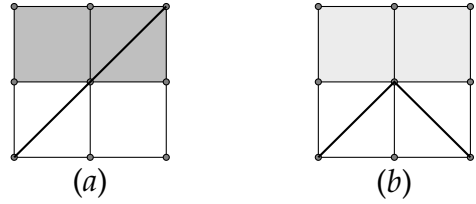

To each point at 
$(i,h_i)$ ($i=1,2,  \cdots,L$), we assign a weight $c_i$ that depends on the shape 
of the path connecting the point with its neighbors and the position of the point. 
When the path enters a ground-state band in the upward direction at point 
$(i,h_i)$ (Figure \ref{fig:weight-1} (a)), or it reaches a peak at $(i,h_i)$ outside 
the ground-state band (Figure \ref{fig:weight-1} (b)), the weight $c_i$ is determined by 

\begin{equation}
    c_i=x_i
\end{equation}

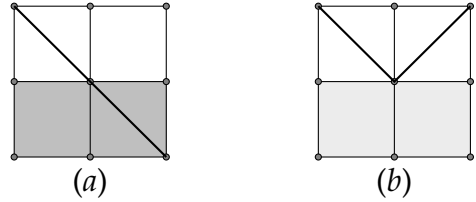
\begin{figure}
    \begin{tikzpicture}
    \draw [thin] (0.0,0,0)--(0.0,2.0);
    \draw [thin] (0.0,0.0)--(2.0,0.0);
    \draw [thin] (0.0,2.0)--(2.0,2.0);
    \draw [thin] (2.0,2.0)--(2.0,0.0);
    \draw [thin] (1.0,0,0)--(1.0,2.0);
    \draw [thin] (0.0,1.0)--(2.0,1.0);
    
    \foreach \x in {0.0,...,2.0}
{
\draw [fill=black!50] (\x,2) circle (0.04);
\draw [fill=black!50] (\x,1) circle (0.04);
\draw [fill=black!50] (\x,0) circle (0.04);
}

\begin{scope}[on background layer]
\fill[gray!50] 
(0.00, 0.00) -- (2.00, 0.00) -- (2.00, 1.00) -- (0.00, 1.00);
\end{scope}

    \draw [thick] (0,2)--(2,0);
    \node [below] at (1,0) {$(a)$};
    
    \draw [thin] (4.0,0,0)--(4.0,2.0);
    \draw [thin] (4.0,0.0)--(6.0,0.0);
    \draw [thin] (4.0,2.0)--(6.0,2.0);
    \draw [thin] (6.0,2.0)--(6.0,0.0);
    \draw [thin] (5.0,0,0)--(5.0,2.0);
    \draw [thin] (4.0,1.0)--(6.0,1.0);
    
    \foreach \x in {4.0,...,6.0}
{
\draw [fill=black!50] (\x,2) circle (0.04);
\draw [fill=black!50] (\x,1) circle (0.04);
\draw [fill=black!50] (\x,0) circle (0.04);
}

\begin{scope}[on background layer]
\fill[gray!15] 
(4.00, 1.00) -- (6.00, 1.00) -- (6.00, 0.00) -- (4.00, 0.00);
\end{scope}

    \draw [thick] (4,2)--(5,1)--(6,2);
    \node [below] at (5,0) {$(b)$};

    \end{tikzpicture}
    \caption{\textit{
    The two cases where we assign the weight $c_i=y_i$ to the middle point
    located at position $i$. Scanning from left to right,  
    (a) the path enters the ground-state (shaded) band from the top at the 
    middle point,  
    (b) the path makes a valley at the middle point in a non-ground-state 
    (white) band. The lightly-shaded band (the lower band in Figure (b)) can be 
    either a ground-state band or not.
    }} 
    \label{fig:weight-2}
\end{figure}

When the path enters in the downward direction to the ground-state band 
(Figure \ref{fig:weight-2} (a)) or hits a valley outside the ground-state band 
(Figure \ref{fig:weight-2} (b)), we assign it 

\begin{equation}
    c_i=y_i
\end{equation}

Otherwise, the weight is set to zero. 
The weight of the whole path is given by the sum of the weights of each point, 
\begin{equation}
    wt(P)=\sum_{i=1}^L c_i
\end{equation}
The \textit{\lq L-finite\rq} (or \textit{\lq finitized\rq}) character for 
a fixed-length lattice and fixed parameters $(a,b,c)$, labeling the start 
and end points, is given by the sum over all allowed finite-length weighted 
paths $P$, 

\begin{equation}
\chi^{\,  p,  \, p'}_{a,b,c}(L,q)=
\sum_{P} \, q^{\, wt(P)}
\end{equation}

The character of the corresponding minimal model is obtained in the limit 
$L\rightarrow \infty$, 

\begin{equation}
\chi^{\,  p,  \, p'}_{r,s=a}(q) = 
\lim_{L \rightarrow \infty} \chi^{\,  p,  \, p'}_{a,b,c} \ll L, q \rr
\end{equation}

\begin{figure}
\begin{tikzpicture}[scale=1.0]

\draw [thin] (0.00, 3.0)--(08.00, 3.0);
\draw [thin] (0.00, 2.0)--(08.00, 2.0);
\draw [thin] (0.00, 1.0)--(08.00, 1.0);
\draw [thin] (0.00, 0.0)--(08.00, 0.0);

\draw [thin] ( 0.0, 0.0)--( 0.0, 3.0);
\draw [thin] ( 1.0, 0.0)--( 1.0, 3.0);
\draw [thin] ( 2.0, 0.0)--( 2.0, 3.0);
\draw [thin] ( 3.0, 0.0)--( 3.0, 3.0);
\draw [thin] ( 4.0, 0.0)--( 4.0, 3.0);
\draw [thin] ( 5.0, 0.0)--( 5.0, 3.0);
\draw [thin] ( 6.0, 0.0)--( 6.0, 3.0);
\draw [thin] ( 7.0, 0.0)--( 7.0, 3.0);
\draw [thin] ( 8.0, 0.0)--( 8.0, 3.0);

\draw [thick] (0.0, 0.0)--(1.0, 1.0);
\draw [thick] (1.0, 1.0)--(2.0, 2.0);
\draw [thick] (2.0, 2.0)--(3.0, 1.0);
\draw [thick] (3.0, 1.0)--(4.0, 2.0);
\draw [thick] (4.0, 2.0)--(5.0, 1.0);
\draw [thick] (5.0, 1.0)--(6.0, 2.0);
\draw [thick] (6.0, 2.0)--(7.0, 1.0);
\draw [thick] (7.0, 1.0)--(8.0, 2.0);

\foreach \x in {0.0,...,8.0}
{
\draw [fill=black!50] (\x,3) circle (0.04);
\draw [fill=black!50] (\x,2) circle (0.04);
\draw [fill=black!50] (\x,1) circle (0.04);
\draw [fill=black!50] (\x,0) circle (0.04);
}
\node at (-0.50, 3.00) {$\textswab{4}$};
\node at (-0.50, 2.00) {$\textswab{3}$};
\node at (-0.50, 1.00) {$\textswab{2}$};
\node at (-0.50, 0.00) {$\textswab{1}$};
\node at ( 8.50, 2.50) {$\textswab{3}$};
\node at ( 8.50, 1.50) {$\textswab{2}$};
\node at ( 8.50, 0.50) {$\textswab{1}$};

\begin{scope}[on background layer]
\fill[gray!50] 
(0.00, 2.00) -- (8.00, 2.00) -- (8.00, 1.00) -- (0.00, 1.00);
\end{scope}

\end{tikzpicture}
\caption{\textit{
A minimal path in the vacuum module of the restricted solid-on-solid 
model $\cL^{\, p, p'} = \cL^{\, 2, 5}$, labeled by $a=1$, $b=2$, and 
$c=3$. 
The numbers on the left label the possible initial points of paths, 
and correspond to the label $s$ in 
$\chi^{\, p, p'}_{\, r, s} = \chi^{\, 2, 5}_{\, r, s}$.
The numbers on the right label the possible locations of 
the ground-state (shaded) bands that paths can end up oscillating in. 
The number of ground-state bands is the range of the label $r$ in 
$\chi^{\, p, p'}_{\, r, s} = \chi^{\, 2, 5}_{\, r, s}$. 
In this example, there is only one ground state band, and
$\chi^{\, p, p'}_{\, r, s} = \chi^{\, 2, 5}_{\, 1, 1}$.
}
}
\label{figure.path.01}
\end{figure}
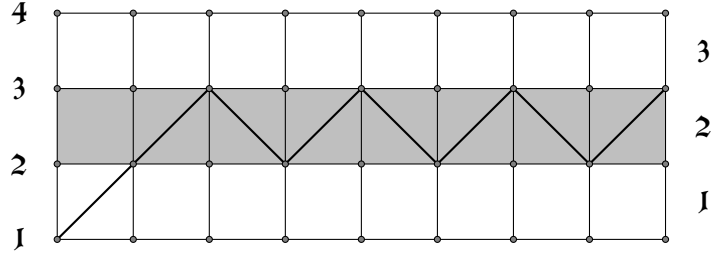

The vacuum module in an $\cL^{\, 2, \, 2 k + 3}$ model is characterized by $r=s=a=1$, 
and $r$ is fixed by $b$ and $c$ through (\ref{r-def}). In principle, there are two 
equivalent combinations
\footnote{\,
Equivalent in the sense that the resulting character is the same.
} 
of $(b,c)$ that give the same value of $r$. We choose the one such that the 
lattice square spanned by the point $(L,b)$ and $(L+1,c)$ is contained inside the 
ground-state band. This fixes $b=2$ and $c=3$ in the Lee-Yang model $\cL^{\, 2, 5}$, as shown in 
Figure \ref{figure.path.01}. More generally, in the case of the models 
$\cL^{\, 2, 2k+3}$, $b=k+1$ and $c=k+2$. 

\subsubsection{Remark}
\textit{
In the case of the models $\cL^{\, 2, \, 2 k + 3}$ (and only 
in this case), one can check that the contributions to the weight of a path come 
effectively from the positions of the peaks and valleys (that is, the positions 
along the horizontal extension of the lattice, which starts from $i=0$), outside 
the ground-state band, and that it costs nothing to wander inside the ground-state 
band. 
A typical path with finite weight (in the $L\rightarrow \infty$ limit) will 
converge into a zigzag inside the ground-state band at finite $i$. 
} 

\subsubsection{Remark}
\textit{
The path with no peaks or valleys outside the ground-state band is the minimal 
path (an example is in Figure \ref{figure.path.01}), and corresponds to the highest 
weight state in the corresponding minimal model module. 
(Virasoro) Descendant states correspond to non-minimal paths.
}

\begin{figure}
\begin{tikzpicture}[scale=1.0]

\draw [thin] (0.00, 3.0)--(08.00, 3.0);
\draw [thin] (0.00, 2.0)--(08.00, 2.0);
\draw [thin] (0.00, 1.0)--(08.00, 1.0);
\draw [thin] (0.00, 0.0)--(08.00, 0.0);

\draw [thin] ( 0.0, 0.0)--( 0.0, 3.0);
\draw [thin] ( 1.0, 0.0)--( 1.0, 3.0);
\draw [thin] ( 2.0, 0.0)--( 2.0, 3.0);
\draw [thin] ( 3.0, 0.0)--( 3.0, 3.0);
\draw [thin] ( 4.0, 0.0)--( 4.0, 3.0);
\draw [thin] ( 5.0, 0.0)--( 5.0, 3.0);
\draw [thin] ( 6.0, 0.0)--( 6.0, 3.0);
\draw [thin] ( 7.0, 0.0)--( 7.0, 3.0);
\draw [thin] ( 8.0, 0.0)--( 8.0, 3.0);

\draw [thick] (0.0, 0.0)--(1.0, 1.0);
\draw [thick] (1.0, 1.0)--(2.0, 0.0);
\draw [thick] (2.0, 0.0)--(3.0, 1.0);
\draw [thick] (3.0, 1.0)--(4.0, 2.0);
\draw [thick] (4.0, 2.0)--(5.0, 3.0);
\draw [thick] (5.0, 3.0)--(6.0, 2.0);
\draw [thick] (6.0, 2.0)--(7.0, 1.0);
\draw [thick] (7.0, 1.0)--(8.0, 2.0);

\foreach \x in {0.0,...,8.0}
{
\draw [fill=black!50] (\x,3) circle (0.04);
\draw [fill=black!50] (\x,2) circle (0.04);
\draw [fill=black!50] (\x,1) circle (0.04);
\draw [fill=black!50] (\x,0) circle (0.04);
}
\node at (-0.50, 3.00)  {$\textswab{4}$};
\node at (-0.50, 2.00)  {$\textswab{3}$};
\node at (-0.50, 1.00)  {$\textswab{2}$};
\node at (-0.50, 0.00)  {$\textswab{1}$};
\node at ( 8.50, 2.50)  {$\textswab{3}$};
\node at ( 8.50, 1.50)  {$\textswab{2}$};
\node at ( 8.50, 0.50)  {$\textswab{1}$};

\begin{scope}[on background layer]
\fill[gray!50] 
(0.00, 2.00) -- (8.00, 2.00) -- (8.00, 1.00) -- (0.00, 1.00);
\end{scope}

\end{tikzpicture}
\caption{\textit{
A non-minimal path in the restricted solid-on-solid model 
$\cL^{\, p, p'} = \cL^{\, 2, 5}$, labeled by $a=1$, $b=2$, 
and $c=3$. 
}}
\label{figure.path.02}
\end{figure}
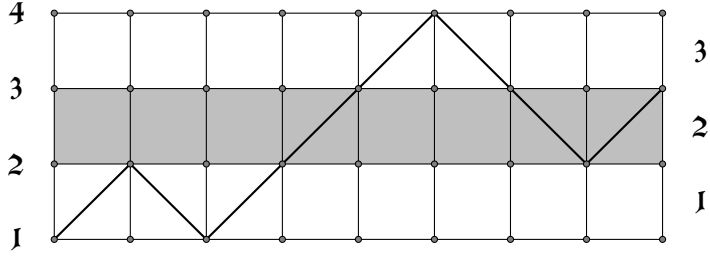

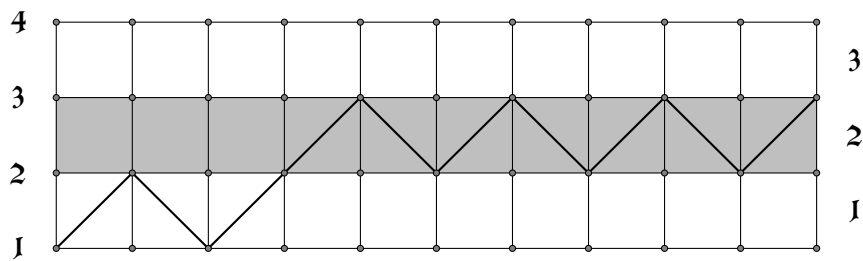
\begin{figure}
\begin{tikzpicture}[scale=1.0]

\draw [thin] (0.00, 3.0)--(10.00, 3.0);
\draw [thin] (0.00, 2.0)--(10.00, 2.0);
\draw [thin] (0.00, 1.0)--(10.00, 1.0);
\draw [thin] (0.00, 0.0)--(10.00, 0.0);

\draw [thin] ( 0.0, 0.0)--( 0.0, 3.0);
\draw [thin] ( 1.0, 0.0)--( 1.0, 3.0);
\draw [thin] ( 2.0, 0.0)--( 2.0, 3.0);
\draw [thin] ( 3.0, 0.0)--( 3.0, 3.0);
\draw [thin] ( 4.0, 0.0)--( 4.0, 3.0);
\draw [thin] ( 5.0, 0.0)--( 5.0, 3.0);
\draw [thin] ( 6.0, 0.0)--( 6.0, 3.0);
\draw [thin] ( 7.0, 0.0)--( 7.0, 3.0);
\draw [thin] ( 8.0, 0.0)--( 8.0, 3.0);
\draw [thin] ( 9.0, 0.0)--( 9.0, 3.0);
\draw [thin] (10.0, 0.0)--(10.0, 3.0);

\draw [thick] (0.0, 0.0)--(1.0, 1.0);
\draw [thick] (1.0, 1.0)--(2.0, 0.0);
\draw [thick] (2.0, 0.0)--(3.0, 1.0);
\draw [thick] (3.0, 1.0)--(4.0, 2.0);
\draw [thick] (4.0, 2.0)--(5.0, 1.0);
\draw [thick] (5.0, 1.0)--(6.0, 2.0);
\draw [thick] (6.0, 2.0)--(7.0, 1.0);
\draw [thick] (7.0, 1.0)--(8.0, 2.0);
\draw [thick] (8.0, 2.0)--(9.0, 1.0);
\draw [thick] (9.0, 1.0)--(10.0, 2.0);

\foreach \x in {0.0,...,10.0}
{
\draw [fill=black!50] (\x,3) circle (0.04);
\draw [fill=black!50] (\x,2) circle (0.04);
\draw [fill=black!50] (\x,1) circle (0.04);
\draw [fill=black!50] (\x,0) circle (0.04);
}
\node at (-0.50, 3.00)  {$\textswab{4}$};
\node at (-0.50, 2.00)  {$\textswab{3}$};
\node at (-0.50, 1.00)  {$\textswab{2}$};
\node at (-0.50, 0.00)  {$\textswab{1}$};
\node at ( 10.50, 2.50)  {$\textswab{3}$};
\node at ( 10.50, 1.50)  {$\textswab{2}$};
\node at ( 10.50, 0.50)  {$\textswab{1}$};

\begin{scope}[on background layer]
\fill[gray!50] 
(0.00, 2.00) -- (10.00, 2.00) -- (10.00, 1.00) -- (0.00, 1.00);
\end{scope}

\end{tikzpicture}
\caption{\textit{
A non-minimal path in the restricted solid-on-solid model $\cL^{\, 2, 5}$, 
labeled by $a=1$, $b=2$, and $c=3$, that contains a valley at $(x,y)=(1,1)$ of weight $wt=1+1=2$. 
}}
\label{figure.valley.02}
\end{figure}

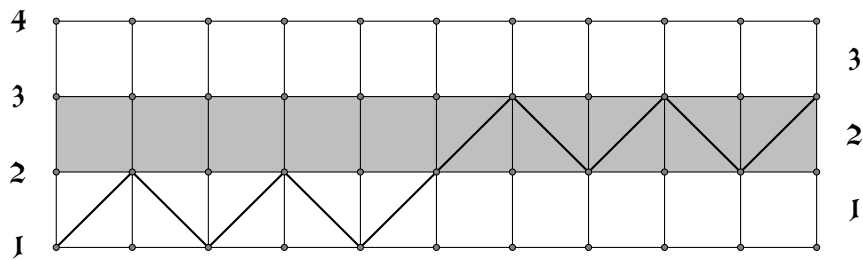
\begin{figure}
\begin{tikzpicture}[scale=1.0]

\draw [thin] (0.00, 3.0)--(10.00, 3.0);
\draw [thin] (0.00, 2.0)--(10.00, 2.0);
\draw [thin] (0.00, 1.0)--(10.00, 1.0);
\draw [thin] (0.00, 0.0)--(10.00, 0.0);

\draw [thin] ( 0.0, 0.0)--( 0.0, 3.0);
\draw [thin] ( 1.0, 0.0)--( 1.0, 3.0);
\draw [thin] ( 2.0, 0.0)--( 2.0, 3.0);
\draw [thin] ( 3.0, 0.0)--( 3.0, 3.0);
\draw [thin] ( 4.0, 0.0)--( 4.0, 3.0);
\draw [thin] ( 5.0, 0.0)--( 5.0, 3.0);
\draw [thin] ( 6.0, 0.0)--( 6.0, 3.0);
\draw [thin] ( 7.0, 0.0)--( 7.0, 3.0);
\draw [thin] ( 8.0, 0.0)--( 8.0, 3.0);
\draw [thin] ( 9.0, 0.0)--( 9.0, 3.0);
\draw [thin] (10.0, 0.0)--(10.0, 3.0);

\draw [thick] (0.0, 0.0)--(1.0, 1.0);
\draw [thick] (1.0, 1.0)--(2.0, 0.0);
\draw [thick] (2.0, 0.0)--(3.0, 1.0);
\draw [thick] (3.0, 1.0)--(4.0, 0.0);
\draw [thick] (4.0, 0.0)--(5.0, 1.0);
\draw [thick] (5.0, 1.0)--(6.0, 2.0);
\draw [thick] (6.0, 2.0)--(7.0, 1.0);
\draw [thick] (7.0, 1.0)--(8.0, 2.0);
\draw [thick] (8.0, 2.0)--(9.0, 1.0);
\draw [thick] (9.0, 1.0)--(10.0, 2.0);

\foreach \x in {0.0,...,10.0}
{
\draw [fill=black!50] (\x,3) circle (0.04);
\draw [fill=black!50] (\x,2) circle (0.04);
\draw [fill=black!50] (\x,1) circle (0.04);
\draw [fill=black!50] (\x,0) circle (0.04);
}
\node at (-0.50, 3.00)  {$\textswab{4}$};
\node at (-0.50, 2.00)  {$\textswab{3}$};
\node at (-0.50, 1.00)  {$\textswab{2}$};
\node at (-0.50, 0.00)  {$\textswab{1}$};
\node at ( 10.50, 2.50)  {$\textswab{3}$};
\node at ( 10.50, 1.50)  {$\textswab{2}$};
\node at ( 10.50, 0.50)  {$\textswab{1}$};

\begin{scope}[on background layer]
\fill[gray!50] 
(0.00, 2.00) -- (10.00, 2.00) -- (10.00, 1.00) -- (0.00, 1.00);
\end{scope}

\end{tikzpicture}
\caption{\textit{
A non-minimal path in the Lee-Yang restricted solid-on-solid model $\cL^{\, 2, 5}$, 
labeled by $a=1$, $b=2$, and $c=3$, that contains a valley at $(x,y)=(1,1)$ 
and another valley at $(2,2)$ of weight $wt=1+1+2+2=6$. 
}}
\label{figure.valley.24}
\end{figure}

An example of a non-minimal path in the Lee-Yang model $\cL^{\, 2, 5}$ is
in Figure \ref{figure.path.02}. 
Each peak and valley outside the ground-state band is assigned a definite weight. 
For example, the path with a single valley of weight $2$ and the path with a single 
valley of weight $4$ in the Lee-Yang model $\cL^{\, 2, 5}$ are shown in 
Figure \ref{figure.valley.02} and \ref{figure.valley.04}, respectively. 
We can also consider a path with both valleys, as in Figure \ref{figure.valley.24}, 
whose weight is $wt = 2 + 4 = 6$. 
The character of the vacuum module for the Lee-Yang model ($k=1$) can then be 
computed as 
\begin{equation}
\sum_{N_1\geq 0}
\sum_{\substack{t_1,t_2,  \cdots,t_{N_1}
\\
t_{i+1}-t_i\geq 2,t_1\geq 2}} \, q^{\, t_1+t_2+  \cdots+t_{N_1}}
=
\sum_{N_1\geq 0}\frac{q^{\, N_1^2+N_1}}{(q)_{N_1}},
\end{equation}

where $N_1$ gives the number of valleys plus peaks, and $t_i$ denotes the corresponding 
weight of the $i$-th valley or peak
\footnote{\,
This computation is explained in detail in \cite{foda.welsh}.
}. 
These peaks and valleys behave as excitations of 
(quasi-)particles, and we refer to them as particles. 

\begin{figure}
\begin{tikzpicture}[scale=1.0]

\draw [thin] (0.00, 3.0)--(10.00, 3.0);
\draw [thin] (0.00, 2.0)--(10.00, 2.0);
\draw [thin] (0.00, 1.0)--(10.00, 1.0);
\draw [thin] (0.00, 0.0)--(10.00, 0.0);

\draw [thin] ( 0.0, 0.0)--( 0.0, 3.0);
\draw [thin] ( 1.0, 0.0)--( 1.0, 3.0);
\draw [thin] ( 2.0, 0.0)--( 2.0, 3.0);
\draw [thin] ( 3.0, 0.0)--( 3.0, 3.0);
\draw [thin] ( 4.0, 0.0)--( 4.0, 3.0);
\draw [thin] ( 5.0, 0.0)--( 5.0, 3.0);
\draw [thin] ( 6.0, 0.0)--( 6.0, 3.0);
\draw [thin] ( 7.0, 0.0)--( 7.0, 3.0);
\draw [thin] ( 8.0, 0.0)--( 8.0, 3.0);
\draw [thin] ( 9.0, 0.0)--( 9.0, 3.0);
\draw [thin] (10.0, 0.0)--(10.0, 3.0);

\draw [thick] (0.0, 0.0)--(1.0, 1.0);
\draw [thick] (1.0, 1.0)--(2.0, 2.0);
\draw [thick] (2.0, 2.0)--(3.0, 1.0);
\draw [thick] (3.0, 1.0)--(4.0, 0.0);
\draw [thick] (4.0, 0.0)--(5.0, 1.0);
\draw [thick] (5.0, 1.0)--(6.0, 2.0);
\draw [thick] (6.0, 2.0)--(7.0, 1.0);
\draw [thick] (7.0, 1.0)--(8.0, 2.0);
\draw [thick] (8.0, 2.0)--(9.0, 1.0);
\draw [thick] (9.0, 1.0)--(10.0, 2.0);

\foreach \x in {0.0,...,10.0}
{
\draw [fill=black!50] (\x,3) circle (0.04);
\draw [fill=black!50] (\x,2) circle (0.04);
\draw [fill=black!50] (\x,1) circle (0.04);
\draw [fill=black!50] (\x,0) circle (0.04);
}
\node at (-0.50, 3.00)  {$\textswab{4}$};
\node at (-0.50, 2.00)  {$\textswab{3}$};
\node at (-0.50, 1.00)  {$\textswab{2}$};
\node at (-0.50, 0.00)  {$\textswab{1}$};
\node at ( 10.50, 2.50)  {$\textswab{3}$};
\node at ( 10.50, 1.50)  {$\textswab{2}$};
\node at ( 10.50, 0.50)  {$\textswab{1}$};

\begin{scope}[on background layer]
\fill[gray!50] 
(0.00, 2.00) -- (10.00, 2.00) -- (10.00, 1.00) -- (0.00, 1.00);
\end{scope}

\end{tikzpicture}
\caption{\textit{
A non-minimal path in the Lee-Yang restricted solid-on-solid model $\cL^{\, 2, 5}$, 
labeled by $a=1$, $b=2$, and $c=3$, that contains a valley at 
$(x,y)=(2,2)$ of weight $wt=2+2=4$. 
}}
\label{figure.valley.04}
\end{figure}

\subsubsection{Higher-$k$ models}

Models with higher $k$ are built using the same rules described in the previous subsection, 
but they are naturally somewhat more complicated. Consider $k=2$, that is $p=2, p'=7$. The lattice 
in Figure \ref{figure.path.03}, of size $6 \times (L+2)$, $L=12$, shows the minimal path in this model. 
For higher $k$, there are $k$ particle species. For $k=2$, there are two different paths, one in 
Figure \ref{figure.path.04} and one in \ref{figure.path.05}, with a single particle each, of 
different particle species, but the same weight, $4$. 
 
\begin{figure}
\begin{tikzpicture}[scale=1.0]

\draw [thin] (0.00, 5.0)--(13.00, 5.0);
\draw [thin] (0.00, 4.0)--(13.00, 4.0);
\draw [thin] (0.00, 3.0)--(13.00, 3.0);
\draw [thin] (0.00, 2.0)--(13.00, 2.0);
\draw [thin] (0.00, 1.0)--(13.00, 1.0);
\draw [thin] (0.00, 0.0)--(13.00, 0.0);

\draw [thin] ( 0.0, 0.0)--( 0.0, 5.0);
\draw [thin] ( 1.0, 0.0)--( 1.0, 5.0);
\draw [thin] ( 2.0, 0.0)--( 2.0, 5.0);
\draw [thin] ( 3.0, 0.0)--( 3.0, 5.0);
\draw [thin] ( 4.0, 0.0)--( 4.0, 5.0);
\draw [thin] ( 5.0, 0.0)--( 5.0, 5.0);
\draw [thin] ( 6.0, 0.0)--( 6.0, 5.0);
\draw [thin] ( 7.0, 0.0)--( 7.0, 5.0);
\draw [thin] ( 8.0, 0.0)--( 8.0, 5.0);
\draw [thin] ( 9.0, 0.0)--( 9.0, 5.0);
\draw [thin] (10.0, 0.0)--(10.0, 5.0);
\draw [thin] (11.0, 0.0)--(11.0, 5.0);
\draw [thin] (12.0, 0.0)--(12.0, 5.0);
\draw [thin] (13.0, 0.0)--(13.0, 5.0);

\draw [thick] (0.0, 0.0)--(1.0, 1.0);
\draw [thick] (1.0, 1.0)--(2.0, 2.0);
\draw [thick] (2.0, 2.0)--(3.0, 3.0);
\draw [thick] (3.0, 3.0)--(4.0, 2.0);
\draw [thick] (4.0, 2.0)--(5.0, 3.0);
\draw [thick] (5.0, 3.0)--(6.0, 2.0);
\draw [thick] (6.0, 2.0)--(7.0, 3.0);
\draw [thick] (7.0, 3.0)--(8.0, 2.0);
\draw [thick] (8.0, 2.0)--(9.0, 3.0);
\draw [thick] (9.0, 3.0)--(10.0, 2.0);
\draw [thick] (10.0, 2.0)--(11.0, 3.0);
\draw [thick] (11.0, 3.0)--(12.0, 2.0);
\draw [thick] (12.0, 2.0)--(13.0, 3.0);

\foreach \x in {0.0,...,13.0}
{
\draw [fill=black!50] (\x,5) circle (0.04);
\draw [fill=black!50] (\x,4) circle (0.04);
\draw [fill=black!50] (\x,3) circle (0.04);
\draw [fill=black!50] (\x,2) circle (0.04);
\draw [fill=black!50] (\x,1) circle (0.04);
\draw [fill=black!50] (\x,0) circle (0.04);
}

\node at (-0.50, 5.00) {$\textswab{6}$};
\node at (-0.50, 4.00) {$\textswab{5}$};
\node at (-0.50, 3.00) {$\textswab{4}$};
\node at (-0.50, 2.00) {$\textswab{3}$};
\node at (-0.50, 1.00) {$\textswab{2}$};
\node at (-0.50, 0.00) {$\textswab{1}$};

\node at (13.50, 4.50) {$\textswab{5}$};
\node at (13.50, 3.50) {$\textswab{4}$};
\node at (13.50, 2.50) {$\textswab{3}$};
\node at (13.50, 1.50) {$\textswab{2}$};
\node at (13.50, 0.50) {$\textswab{1}$};

\begin{scope}[on background layer]
\fill[gray!50] 
(0.00, 3.00) -- (13.00, 3.00) -- (13.00, 2.00) -- (0.00, 2.00);
\end{scope}

\end{tikzpicture}
\caption{\textit{
The minimal path in the vacuum module of the restricted solid-on-solid model 
$\cL^{\, 2, 7}$, labeled by $a=1$, $b=3$, and $c=4$. The weight of a minimal 
path is zero. 
}}
\label{figure.path.03}
\end{figure}
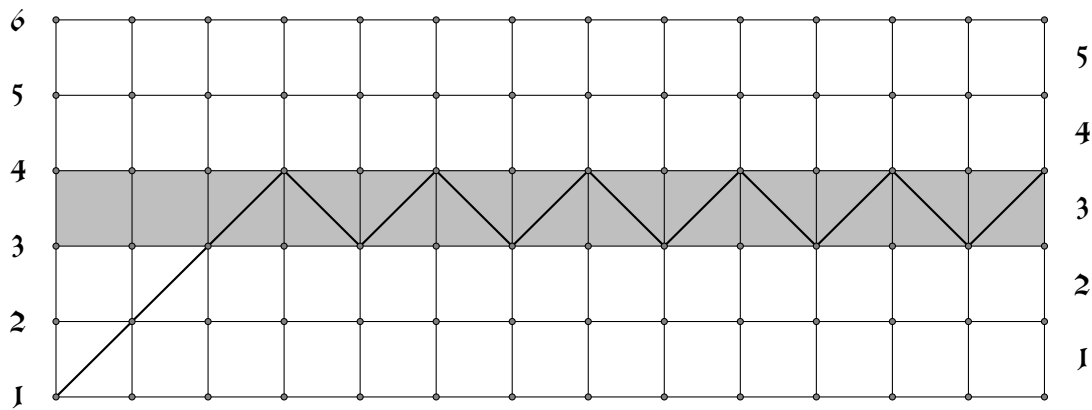

\begin{figure}
\begin{tikzpicture}[scale=1.0]

\draw [thin] (0.00, 5.0)--(13.00, 5.0);
\draw [thin] (0.00, 4.0)--(13.00, 4.0);
\draw [thin] (0.00, 3.0)--(13.00, 3.0);
\draw [thin] (0.00, 2.0)--(13.00, 2.0);
\draw [thin] (0.00, 1.0)--(13.00, 1.0);
\draw [thin] (0.00, 0.0)--(13.00, 0.0);

\draw [thin] ( 0.0, 0.0)--( 0.0, 5.0);
\draw [thin] ( 1.0, 0.0)--( 1.0, 5.0);
\draw [thin] ( 2.0, 0.0)--( 2.0, 5.0);
\draw [thin] ( 3.0, 0.0)--( 3.0, 5.0);
\draw [thin] ( 4.0, 0.0)--( 4.0, 5.0);
\draw [thin] ( 5.0, 0.0)--( 5.0, 5.0);
\draw [thin] ( 6.0, 0.0)--( 6.0, 5.0);
\draw [thin] ( 7.0, 0.0)--( 7.0, 5.0);
\draw [thin] ( 8.0, 0.0)--( 8.0, 5.0);
\draw [thin] ( 9.0, 0.0)--( 9.0, 5.0);
\draw [thin] (10.0, 0.0)--(10.0, 5.0);
\draw [thin] (11.0, 0.0)--(11.0, 5.0);
\draw [thin] (12.0, 0.0)--(12.0, 5.0);
\draw [thin] (13.0, 0.0)--(13.0, 5.0);

\draw [thick] (0.0, 0.0)--(1.0, 1.0);
\draw [thick] (1.0, 1.0)--(2.0, 2.0);
\draw [thick] (2.0, 2.0)--(3.0, 3.0);
\draw [thick] (3.0, 3.0)--(4.0, 4.0);
\draw [thick] (4.0, 4.0)--(5.0, 3.0);
\draw [thick] (5.0, 3.0)--(6.0, 2.0);
\draw [thick] (6.0, 2.0)--(7.0, 3.0);
\draw [thick] (7.0, 3.0)--(8.0, 2.0);
\draw [thick] (8.0, 2.0)--(9.0, 3.0);
\draw [thick] (9.0, 3.0)--(10.0, 2.0);
\draw [thick] (10.0, 2.0)--(11.0, 3.0);
\draw [thick] (11.0, 3.0)--(12.0, 2.0);
\draw [thick] (12.0, 2.0)--(13.0, 3.0);

\foreach \x in {0.0,...,13.0}
{
\draw [fill=black!50] (\x,5) circle (0.04);
\draw [fill=black!50] (\x,4) circle (0.04);
\draw [fill=black!50] (\x,3) circle (0.04);
\draw [fill=black!50] (\x,2) circle (0.04);
\draw [fill=black!50] (\x,1) circle (0.04);
\draw [fill=black!50] (\x,0) circle (0.04);
}

\node at (-0.50, 5.00) {$\textswab{6}$};
\node at (-0.50, 4.00) {$\textswab{5}$};
\node at (-0.50, 3.00) {$\textswab{4}$};
\node at (-0.50, 2.00) {$\textswab{3}$};
\node at (-0.50, 1.00) {$\textswab{2}$};
\node at (-0.50, 0.00) {$\textswab{1}$};

\node at (13.50, 4.50) {$\textswab{5}$};
\node at (13.50, 3.50) {$\textswab{4}$};
\node at (13.50, 2.50) {$\textswab{3}$};
\node at (13.50, 1.50) {$\textswab{2}$};
\node at (13.50, 0.50) {$\textswab{1}$};

\begin{scope}[on background layer]
\fill[gray!50] 
(0.00, 3.00) -- (13.00, 3.00) -- (13.00, 2.00) -- (0.00, 2.00);
\end{scope}

\end{tikzpicture}
\caption{\textit{
A non-minimal path in the restricted solid-on-solid model $\cL^{\, 2, 7}$ 
with a peak at $(x,y)=(0,4)$ of weight $4$. 
}}
\label{figure.path.04}
\end{figure}

\begin{figure}
\begin{tikzpicture}[scale=1.0]

\draw [thin] (0.00, 5.0)--(13.00, 5.0);
\draw [thin] (0.00, 4.0)--(13.00, 4.0);
\draw [thin] (0.00, 3.0)--(13.00, 3.0);
\draw [thin] (0.00, 2.0)--(13.00, 2.0);
\draw [thin] (0.00, 1.0)--(13.00, 1.0);
\draw [thin] (0.00, 0.0)--(13.00, 0.0);

\draw [thin] ( 0.0, 0.0)--( 0.0, 5.0);
\draw [thin] ( 1.0, 0.0)--( 1.0, 5.0);
\draw [thin] ( 2.0, 0.0)--( 2.0, 5.0);
\draw [thin] ( 3.0, 0.0)--( 3.0, 5.0);
\draw [thin] ( 4.0, 0.0)--( 4.0, 5.0);
\draw [thin] ( 5.0, 0.0)--( 5.0, 5.0);
\draw [thin] ( 6.0, 0.0)--( 6.0, 5.0);
\draw [thin] ( 7.0, 0.0)--( 7.0, 5.0);
\draw [thin] ( 8.0, 0.0)--( 8.0, 5.0);
\draw [thin] ( 9.0, 0.0)--( 9.0, 5.0);
\draw [thin] (10.0, 0.0)--(10.0, 5.0);
\draw [thin] (11.0, 0.0)--(11.0, 5.0);
\draw [thin] (12.0, 0.0)--(12.0, 5.0);
\draw [thin] (13.0, 0.0)--(13.0, 5.0);

\draw [thick] (0.0, 0.0)--(1.0, 1.0);
\draw [thick] (1.0, 1.0)--(2.0, 2.0);
\draw [thick] (2.0, 2.0)--(3.0, 1.0);
\draw [thick] (3.0, 1.0)--(4.0, 0.0);
\draw [thick] (4.0, 0.0)--(5.0, 1.0);
\draw [thick] (5.0, 1.0)--(6.0, 2.0);
\draw [thick] (6.0, 2.0)--(7.0, 3.0);
\draw [thick] (7.0, 3.0)--(8.0, 2.0);
\draw [thick] (8.0, 2.0)--(9.0, 3.0);
\draw [thick] (9.0, 3.0)--(10.0, 2.0);
\draw [thick] (10.0, 2.0)--(11.0, 3.0);
\draw [thick] (11.0, 3.0)--(12.0, 2.0);
\draw [thick] (12.0, 2.0)--(13.0, 3.0);

\foreach \x in {0.0,...,13.0}
{
\draw [fill=black!50] (\x,5) circle (0.04);
\draw [fill=black!50] (\x,4) circle (0.04);
\draw [fill=black!50] (\x,3) circle (0.04);
\draw [fill=black!50] (\x,2) circle (0.04);
\draw [fill=black!50] (\x,1) circle (0.04);
\draw [fill=black!50] (\x,0) circle (0.04);
}

\node at (-0.50, 5.00) {$\textswab{6}$};
\node at (-0.50, 4.00) {$\textswab{5}$};
\node at (-0.50, 3.00) {$\textswab{4}$};
\node at (-0.50, 2.00) {$\textswab{3}$};
\node at (-0.50, 1.00) {$\textswab{2}$};
\node at (-0.50, 0.00) {$\textswab{1}$};

\node at (13.50, 4.50) {$\textswab{5}$};
\node at (13.50, 3.50) {$\textswab{4}$};
\node at (13.50, 2.50) {$\textswab{3}$};
\node at (13.50, 1.50) {$\textswab{2}$};
\node at (13.50, 0.50) {$\textswab{1}$};

\begin{scope}[on background layer]
\fill[gray!50] 
(0.00, 3.00) -- (13.00, 3.00) -- (13.00, 2.00) -- (0.00, 2.00);
\end{scope}

\end{tikzpicture}
\caption{\textit{
A non-minimal path in the restricted solid-on-solid model $\cL^{\, 2, 7}$, 
with a valley at $(x,y)=(2,2)$ of weight $4$. The valley between the $3$-rd vertical 
line (at $i=2$) and the $7$-th vertical line (at $i=6$) is obtained by reflecting 
with respect to the lower boundary of the ground-state band from the peak in Figure \ref{figure.path.04}. 
}}
\label{figure.path.05}
\end{figure}
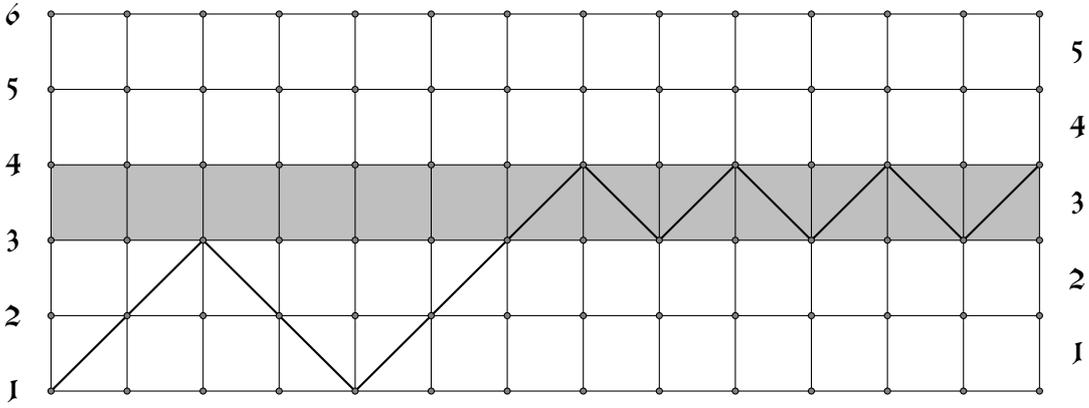

\subsubsection{Remark} 
\textit{We note that one can judge the type of a given particle by using the moves defined in \cite{foda.lee.pugai.welsh} to transform it to the particle with minimal weight of the same type.
}

\subsubsection{Remark} 
\begin{figure}
\begin{tikzpicture}[scale=1.0]

\draw [thin] (0.00, 5.0)--(13.00, 5.0);
\draw [thin] (0.00, 4.0)--(13.00, 4.0);
\draw [thin] (0.00, 3.0)--(13.00, 3.0);
\draw [thin] (0.00, 2.0)--(13.00, 2.0);
\draw [thin] (0.00, 1.0)--(13.00, 1.0);
\draw [thin] (0.00, 0.0)--(13.00, 0.0);

\draw [thin] ( 0.0, 0.0)--( 0.0, 5.0);
\draw [thin] ( 1.0, 0.0)--( 1.0, 5.0);
\draw [thin] ( 2.0, 0.0)--( 2.0, 5.0);
\draw [thin] ( 3.0, 0.0)--( 3.0, 5.0);
\draw [thin] ( 4.0, 0.0)--( 4.0, 5.0);
\draw [thin] ( 5.0, 0.0)--( 5.0, 5.0);
\draw [thin] ( 6.0, 0.0)--( 6.0, 5.0);
\draw [thin] ( 7.0, 0.0)--( 7.0, 5.0);
\draw [thin] ( 8.0, 0.0)--( 8.0, 5.0);
\draw [thin] ( 9.0, 0.0)--( 9.0, 5.0);
\draw [thin] (10.0, 0.0)--(10.0, 5.0);
\draw [thin] (11.0, 0.0)--(11.0, 5.0);
\draw [thin] (12.0, 0.0)--(12.0, 5.0);
\draw [thin] (13.0, 0.0)--(13.0, 5.0);

\draw [thick] (0.0, 0.0)--(1.0, 1.0);
\draw [thick] (1.0, 1.0)--(2.0, 2.0);
\draw [thick] (2.0, 2.0)--(3.0, 3.0);
\draw [thick] (3.0, 3.0)--(4.0, 4.0);
\draw [thick] (4.0, 4.0)--(5.0, 5.0);
\draw [thick] (5.0, 5.0)--(6.0, 4.0);
\draw [thick] (6.0, 4.0)--(7.0, 3.0);
\draw [thick] (7.0, 3.0)--(8.0, 2.0);
\draw [thick] (8.0, 2.0)--(9.0, 3.0);
\draw [thick] (9.0, 3.0)--(10.0, 2.0);
\draw [thick] (10.0, 2.0)--(11.0, 3.0);
\draw [thick] (11.0, 3.0)--(12.0, 2.0);
\draw [thick] (12.0, 2.0)--(13.0, 3.0);

\foreach \x in {0.0,...,13.0}
{
\draw [fill=black!50] (\x,5) circle (0.04);
\draw [fill=black!50] (\x,4) circle (0.04);
\draw [fill=black!50] (\x,3) circle (0.04);
\draw [fill=black!50] (\x,2) circle (0.04);
\draw [fill=black!50] (\x,1) circle (0.04);
\draw [fill=black!50] (\x,0) circle (0.04);
}

\node at (-0.50, 5.00) {$\textswab{6}$};
\node at (-0.50, 4.00) {$\textswab{5}$};
\node at (-0.50, 3.00) {$\textswab{4}$};
\node at (-0.50, 2.00) {$\textswab{3}$};
\node at (-0.50, 1.00) {$\textswab{2}$};
\node at (-0.50, 0.00) {$\textswab{1}$};

\node at (13.50, 4.50) {$\textswab{5}$};
\node at (13.50, 3.50) {$\textswab{4}$};
\node at (13.50, 2.50) {$\textswab{3}$};
\node at (13.50, 1.50) {$\textswab{2}$};
\node at (13.50, 0.50) {$\textswab{1}$};

\begin{scope}[on background layer]
\fill[gray!50] 
(0.00, 3.00) -- (13.00, 3.00) -- (13.00, 2.00) -- (0.00, 2.00);
\end{scope}

\end{tikzpicture}
\caption{\textit{
A non-minimal path in the restricted solid-on-solid model $\cL^{\, 2, 7}$ with 
a peak at $(x,y)=(0,5)$ of weight $5$. 
}}
\label{figure.path.06}
\end{figure}
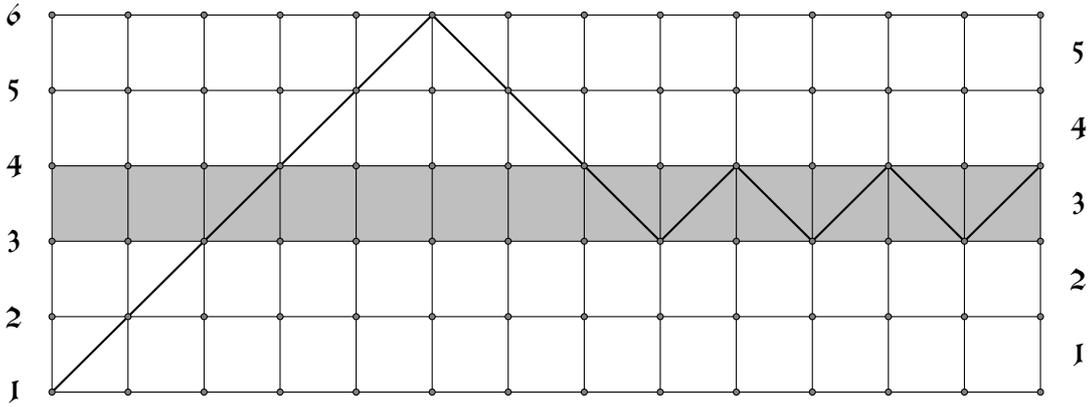

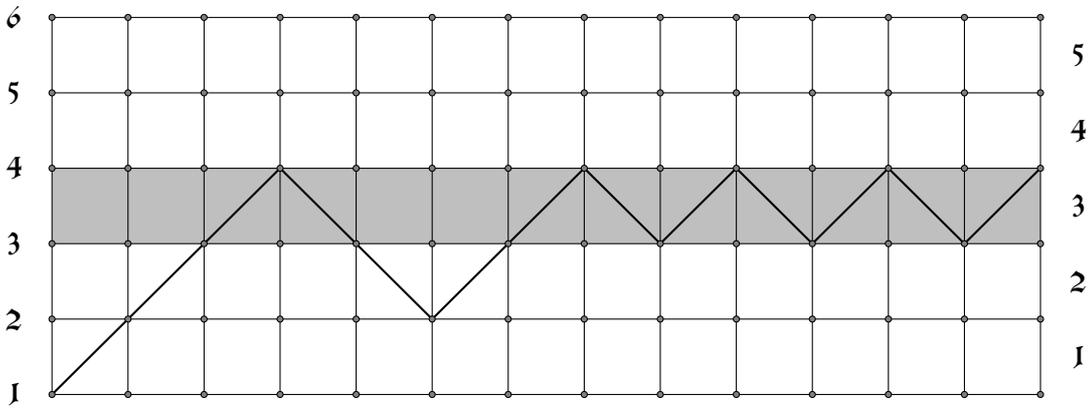
\begin{figure}
\begin{tikzpicture}[scale=1.0]

\draw [thin] (0.00, 5.0)--(13.00, 5.0);
\draw [thin] (0.00, 4.0)--(13.00, 4.0);
\draw [thin] (0.00, 3.0)--(13.00, 3.0);
\draw [thin] (0.00, 2.0)--(13.00, 2.0);
\draw [thin] (0.00, 1.0)--(13.00, 1.0);
\draw [thin] (0.00, 0.0)--(13.00, 0.0);

\draw [thin] ( 0.0, 0.0)--( 0.0, 5.0);
\draw [thin] ( 1.0, 0.0)--( 1.0, 5.0);
\draw [thin] ( 2.0, 0.0)--( 2.0, 5.0);
\draw [thin] ( 3.0, 0.0)--( 3.0, 5.0);
\draw [thin] ( 4.0, 0.0)--( 4.0, 5.0);
\draw [thin] ( 5.0, 0.0)--( 5.0, 5.0);
\draw [thin] ( 6.0, 0.0)--( 6.0, 5.0);
\draw [thin] ( 7.0, 0.0)--( 7.0, 5.0);
\draw [thin] ( 8.0, 0.0)--( 8.0, 5.0);
\draw [thin] ( 9.0, 0.0)--( 9.0, 5.0);
\draw [thin] (10.0, 0.0)--(10.0, 5.0);
\draw [thin] (11.0, 0.0)--(11.0, 5.0);
\draw [thin] (12.0, 0.0)--(12.0, 5.0);
\draw [thin] (13.0, 0.0)--(13.0, 5.0);

\draw [thick] (0.0, 0.0)--(1.0, 1.0);
\draw [thick] (1.0, 1.0)--(2.0, 2.0);
\draw [thick] (2.0, 2.0)--(3.0, 3.0);
\draw [thick] (3.0, 3.0)--(4.0, 2.0);
\draw [thick] (4.0, 2.0)--(5.0, 1.0);
\draw [thick] (5.0, 1.0)--(6.0, 2.0);
\draw [thick] (6.0, 2.0)--(7.0, 3.0);
\draw [thick] (7.0, 3.0)--(8.0, 2.0);
\draw [thick] (8.0, 2.0)--(9.0, 3.0);
\draw [thick] (9.0, 3.0)--(10.0, 2.0);
\draw [thick] (10.0, 2.0)--(11.0, 3.0);
\draw [thick] (11.0, 3.0)--(12.0, 2.0);
\draw [thick] (12.0, 2.0)--(13.0, 3.0);

\foreach \x in {0.0,...,13.0}
{
\draw [fill=black!50] (\x,5) circle (0.04);
\draw [fill=black!50] (\x,4) circle (0.04);
\draw [fill=black!50] (\x,3) circle (0.04);
\draw [fill=black!50] (\x,2) circle (0.04);
\draw [fill=black!50] (\x,1) circle (0.04);
\draw [fill=black!50] (\x,0) circle (0.04);
}

\node at (-0.50, 5.00) {$\textswab{6}$};
\node at (-0.50, 4.00) {$\textswab{5}$};
\node at (-0.50, 3.00) {$\textswab{4}$};
\node at (-0.50, 2.00) {$\textswab{3}$};
\node at (-0.50, 1.00) {$\textswab{2}$};
\node at (-0.50, 0.00) {$\textswab{1}$};

\node at (13.50, 4.50) {$\textswab{5}$};
\node at (13.50, 3.50) {$\textswab{4}$};
\node at (13.50, 2.50) {$\textswab{3}$};
\node at (13.50, 1.50) {$\textswab{2}$};
\node at (13.50, 0.50) {$\textswab{1}$};

\begin{scope}[on background layer]
\fill[gray!50] 
(0.00, 3.00) -- (13.00, 3.00) -- (13.00, 2.00) -- (0.00, 2.00);
\end{scope}

\end{tikzpicture}
\caption{\textit{
A non-minimal path in the restricted solid-on-solid model $\cL_{\, 2, 7}$. 
The valley between 
the $4$-th vertical line (position $i=3$) and 
the $8$-th vertical line (position $i=7$) 
is obtained by reflecting the peak in Figure \ref{figure.path.06} with 
respect to the upper boundary of the ground-state band. 
}}
\label{figure.path.07}
\end{figure}

\textit{
For $k=2, 3, \cdots$, there is a $\mathbb{Z}_2$ reflection 
symmetry between the peaks and valleys with the same weight.  
This symmetry is clear when we compare 
Figures \ref{figure.path.04} and \ref{figure.path.05} 
with weight-$4$, and the reflection symmetry is with respect to the lower boundary of 
the ground-state band. Similarly, when we compare 
Figure \ref{figure.path.06} and \ref{figure.path.07},
with weight-$5$, and the reflection symmetry is with respect to the upper boundary of 
the ground-state band. For $k>2$, we have more than two types of particles, and there will be a $\mathbb{Z}_2$ reflection symmetry between each pair of two different species. We will interpret these quasi-particles as BPS operators in the context of gauge theory, however, it is not clear what kind of role these $\mathbb{Z}_2$ reflection symmetries play there. 
}

The constant-sign sum expression of the $k=2$ vacuum character is

\begin{equation}
\sum_{N_1\geq N_2\geq0} 
\frac{q^{\, N_1^2+N_2^2+N_1+N_2}}{(q)_{N_1-N_2}(q)_{N_2}}
=
1 + \frac{q^2}{1-q} + \frac{q^4}{1-q} + \frac{q^6}{(1-q)(1-q^2)} + \frac{q^8}{(1-q)^2} +  
\cdots
\end{equation}

The term $\frac{q^2}{1-q}$ represents the contributions from all paths with a single 
valley (such as the paths in Figure \ref{figure.path.05} and \ref{figure.path.07}). 
The term $\frac{q^4}{1-q}$, however, comes from the contributions of all paths with 
a single peak of weight larger than $3$ (such as the paths in Figure \ref{figure.path.04} 
and \ref{figure.path.06}). The term $\frac{q^6}{(1-q)(1-q^2)}$ and $\frac{q^8}{(1-q)^2}$ 
can thus be interpreted respectively as the contributions from paths with two valleys 
and paths with one peak and one valley. In this way, we see that $N_1$ in this example 
counts the number of all particles, while $N_2$ counts the number of particles of the 
same type as those in Figure \ref{figure.path.04} and \ref{figure.path.06}, that's is
particles of height 1 above the ground-state band. 

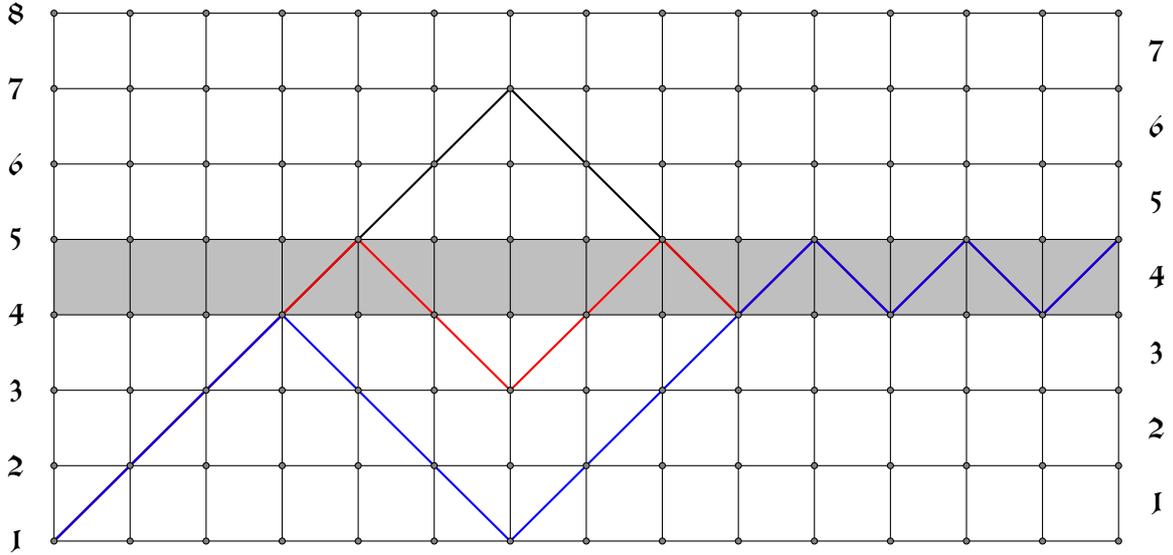
\begin{figure}
\begin{tikzpicture}[scale=1.0]

\draw [thin] (0.00, 7.0)--(14.00, 7.0);
\draw [thin] (0.00, 6.0)--(14.00, 6.0);
\draw [thin] (0.00, 5.0)--(14.00, 5.0);
\draw [thin] (0.00, 4.0)--(14.00, 4.0);
\draw [thin] (0.00, 3.0)--(14.00, 3.0);
\draw [thin] (0.00, 2.0)--(14.00, 2.0);
\draw [thin] (0.00, 1.0)--(14.00, 1.0);
\draw [thin] (0.00, 0.0)--(14.00, 0.0);

\draw [thin] ( 0.0, 0.0)--( 0.0, 7.0);
\draw [thin] ( 1.0, 0.0)--( 1.0, 7.0);
\draw [thin] ( 2.0, 0.0)--( 2.0, 7.0);
\draw [thin] ( 3.0, 0.0)--( 3.0, 7.0);
\draw [thin] ( 4.0, 0.0)--( 4.0, 7.0);
\draw [thin] ( 5.0, 0.0)--( 5.0, 7.0);
\draw [thin] ( 6.0, 0.0)--( 6.0, 7.0);
\draw [thin] ( 7.0, 0.0)--( 7.0, 7.0);
\draw [thin] ( 8.0, 0.0)--( 8.0, 7.0);
\draw [thin] ( 9.0, 0.0)--( 9.0, 7.0);
\draw [thin] (10.0, 0.0)--(10.0, 7.0);
\draw [thin] (11.0, 0.0)--(11.0, 7.0);
\draw [thin] (12.0, 0.0)--(12.0, 7.0);
\draw [thin] (13.0, 0.0)--(13.0, 7.0);
\draw [thin] (14.0, 0.0)--(14.0, 7.0);

\draw [thick] (0.0, 0.0)--(1.0, 1.0);
\draw [thick] (1.0, 1.0)--(2.0, 2.0);
\draw [thick] (2.0, 2.0)--(3.0, 3.0);
\draw [thick] (3.0, 3.0)--(4.0, 4.0);
\draw [thick] (4.0, 4.0)--(5.0, 5.0);
\draw [thick] (5.0, 5.0)--(6.0, 6.0);
\draw [thick] (6.0, 6.0)--(7.0, 5.0);
\draw [thick] (7.0, 5.0)--(8.0, 4.0);
\draw [thick] (8.0, 4.0)--(9.0, 3.0);
\draw [thick] (9.0, 3.0)--(10.0, 4.0);
\draw [thick] (10.0, 4.0)--(11.0, 3.0);
\draw [thick] (11.0, 3.0)--(12.0, 4.0);
\draw [thick] (12.0, 4.0)--(13.0, 3.0);
\draw [thick] (13.0, 3.0)--(14.0, 4.0);

\draw [thick,red] (0.0, 0.0)--(1.0, 1.0);
\draw [thick,red] (1.0, 1.0)--(2.0, 2.0);
\draw [thick,red] (2.0, 2.0)--(3.0, 3.0);
\draw [thick,red] (3.0, 3.0)--(4.0, 4.0);
\draw [thick,red] (4.0, 4.0)--(5.0, 3.0);
\draw [thick,red] (5.0, 3.0)--(6.0, 2.0);
\draw [thick,red] (6.0, 2.0)--(7.0, 3.0);
\draw [thick,red] (7.0, 3.0)--(8.0, 4.0);
\draw [thick,red] (8.0, 4.0)--(9.0, 3.0);
\draw [thick,red] (9.0, 3.0)--(10.0, 4.0);
\draw [thick,red] (10.0, 4.0)--(11.0, 3.0);
\draw [thick,red] (11.0, 3.0)--(12.0, 4.0);
\draw [thick,red] (12.0, 4.0)--(13.0, 3.0);
\draw [thick,red] (13.0, 3.0)--(14.0, 4.0);

\draw [thick,blue] (0.0, 0.0)--(1.0, 1.0);
\draw [thick,blue] (1.0, 1.0)--(2.0, 2.0);
\draw [thick,blue] (2.0, 2.0)--(3.0, 3.0);
\draw [thick,blue] (3.0, 3.0)--(4.0, 2.0);
\draw [thick,blue] (4.0, 2.0)--(5.0, 1.0);
\draw [thick,blue] (5.0, 1.0)--(6.0, 0.0);
\draw [thick,blue] (6.0, 0.0)--(7.0, 1.0);
\draw [thick,blue] (7.0, 1.0)--(8.0, 2.0);
\draw [thick,blue] (8.0, 2.0)--(9.0, 3.0);
\draw [thick,blue] (9.0, 3.0)--(10.0, 4.0);
\draw [thick,blue] (10.0, 4.0)--(11.0, 3.0);
\draw [thick,blue] (11.0, 3.0)--(12.0, 4.0);
\draw [thick,blue] (12.0, 4.0)--(13.0, 3.0);
\draw [thick,blue] (13.0, 3.0)--(14.0, 4.0);

\foreach \x in {0.0,...,14.0}
{
\draw [fill=black!50] (\x,7) circle (0.04);
\draw [fill=black!50] (\x,6) circle (0.04);
\draw [fill=black!50] (\x,5) circle (0.04);
\draw [fill=black!50] (\x,4) circle (0.04);
\draw [fill=black!50] (\x,3) circle (0.04);
\draw [fill=black!50] (\x,2) circle (0.04);
\draw [fill=black!50] (\x,1) circle (0.04);
\draw [fill=black!50] (\x,0) circle (0.04);
}

\node at (-0.50, 7.00) {$\textswab{8}$};
\node at (-0.50, 6.00) {$\textswab{7}$};
\node at (-0.50, 5.00) {$\textswab{6}$};
\node at (-0.50, 4.00) {$\textswab{5}$};
\node at (-0.50, 3.00) {$\textswab{4}$};
\node at (-0.50, 2.00) {$\textswab{3}$};
\node at (-0.50, 1.00) {$\textswab{2}$};
\node at (-0.50, 0.00) {$\textswab{1}$};

\node at (14.50, 6.50) {$\textswab{7}$};
\node at (14.50, 5.50) {$\textswab{6}$};
\node at (14.50, 4.50) {$\textswab{5}$};
\node at (14.50, 3.50) {$\textswab{4}$};
\node at (14.50, 2.50) {$\textswab{3}$};
\node at (14.50, 1.50) {$\textswab{2}$};
\node at (14.50, 0.50) {$\textswab{1}$};

\begin{scope}[on background layer]
\fill[gray!50] 
(0.00, 4.00) -- (14.00, 4.00) -- (14.00, 3.00) -- (0.00, 3.00);
\end{scope}

\end{tikzpicture}
\caption{\textit{
Three paths with single peak/valley in the restricted solid-on-solid model 
$\cL^{\, 2, 9}$ respectively shown in black, red and blue colors.  
}}
\label{figure.various.peaks}
\end{figure}

In the $\cL^{\, 2, 2k+3}$ model, we can have $k$ types of peaks/valleys with 
the same weight. $N_1$ always counts the total number of all particles, and 
$N_i$, $i > 1$ counts the number of different particle species. For example, 
the three paths with weight $6$, in the case $k=3$, are shown in different 
colors in Figure \ref{figure.various.peaks}. 

\subsection{The paths of constant-sign Virasoro characters. The next-to-vacuum modules}

\begin{figure}
\begin{tikzpicture}[scale=1.0]

\draw [thin] (0.00, 3.0)--(09.00, 3.0);
\draw [thin] (0.00, 2.0)--(09.00, 2.0);
\draw [thin] (0.00, 1.0)--(09.00, 1.0);
\draw [thin] (0.00, 0.0)--(09.00, 0.0);

\draw [thin] ( 0.0, 0.0)--( 0.0, 3.0);
\draw [thin] ( 1.0, 0.0)--( 1.0, 3.0);
\draw [thin] ( 2.0, 0.0)--( 2.0, 3.0);
\draw [thin] ( 3.0, 0.0)--( 3.0, 3.0);
\draw [thin] ( 4.0, 0.0)--( 4.0, 3.0);
\draw [thin] ( 5.0, 0.0)--( 5.0, 3.0);
\draw [thin] ( 6.0, 0.0)--( 6.0, 3.0);
\draw [thin] ( 7.0, 0.0)--( 7.0, 3.0);
\draw [thin] ( 8.0, 0.0)--( 8.0, 3.0);
\draw [thin] ( 9.0, 0.0)--( 9.0, 3.0);

\draw [thick] (0.0, 1.0)--(1.0, 2.0);
\draw [thick] (1.0, 2.0)--(2.0, 1.0);
\draw [thick] (2.0, 1.0)--(3.0, 2.0);
\draw [thick] (3.0, 2.0)--(4.0, 1.0);
\draw [thick] (4.0, 1.0)--(5.0, 2.0);
\draw [thick] (5.0, 2.0)--(6.0, 1.0);
\draw [thick] (6.0, 1.0)--(7.0, 2.0);
\draw [thick] (7.0, 2.0)--(8.0, 1.0);
\draw [thick] (8.0, 1.0)--(9.0, 2.0);

\draw [thick,red] (0.0, 1.0)--(1.0, 0.0);
\draw [thick,red] (1.0, 0.0)--(2.0, 1.0);
\draw [thick,red] (2.0, 1.0)--(3.0, 2.0);
\draw [thick,red] (3.0, 2.0)--(4.0, 1.0);
\draw [thick,red] (4.0, 1.0)--(5.0, 2.0);
\draw [thick,red] (5.0, 2.0)--(6.0, 1.0);
\draw [thick,red] (6.0, 1.0)--(7.0, 2.0);
\draw [thick,red] (7.0, 2.0)--(8.0, 1.0);
\draw [thick,red] (8.0, 1.0)--(9.0, 2.0);

\foreach \x in {0.0,...,9.0}
{
\draw [fill=black!50] (\x,3) circle (0.04);
\draw [fill=black!50] (\x,2) circle (0.04);
\draw [fill=black!50] (\x,1) circle (0.04);
\draw [fill=black!50] (\x,0) circle (0.04);
}
\node at (-0.50, 3.00) {$\textswab{4}$};
\node at (-0.50, 2.00) {$\textswab{3}$};
\node at (-0.50, 1.00) {$\textswab{2}$};
\node at (-0.50, 0.00) {$\textswab{1}$};
\node at ( 9.50, 2.50) {$\textswab{3}$};
\node at ( 9.50, 1.50) {$\textswab{2}$};
\node at ( 9.50, 0.50) {$\textswab{1}$};

\begin{scope}[on background layer]
\fill[gray!50] 
(0.00, 2.00) -- (9.00, 2.00) -- (9.00, 1.00) -- (0.00, 1.00);
\end{scope}

\end{tikzpicture}
\caption{\textit{
The minimal path (black) and a path with weight $1$ (red) in the next-to-vacuum module of the Lee-Yang restricted solid-on-solid model 
$\cL^{\, 2, 5}$, labeled by $a=2$, $b=2$, and $c=3$. 
}}
\label{figure.path.01.next}
\end{figure}
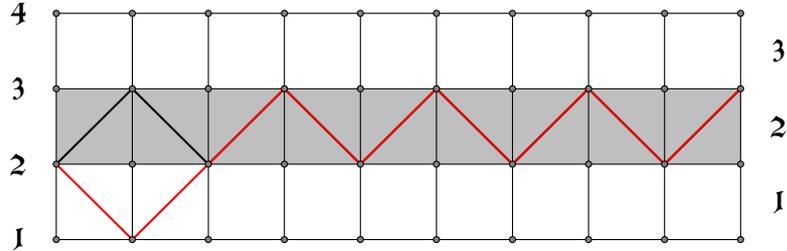

To go to non-vacuum modules, we change $a$ to values larger than $1$. 
The next-to-vacuum module corresponds to $a=2$. 
For example in the Lee-Yang model $\cL^{\, 2, 5}$, the corresponding 
primary Virasoro field has conformal dimension $\Delta=-\frac{1}{5}$. 
The minimal path (with weight zero) 
that corresponds to the highest weight state is shown as the black line in 
Figure \ref{figure.path.01.next}. A direct consequence of the changing value of $a$ is 
is the appearance of new particle configurations with weight $1$ (see the red path shown 
in Figure \ref{figure.path.01.next}). The character of this module is then modified to 

\begin{equation}
\sum_{N_1\geq 0}
\sum_{\substack{t_1,t_2, \cdots, t_{N_1}
\\
t_{i + 1} - t_i \geq 2, t_1 \geq 1}} \, q^{\, t_1 + t_2 + \cdots + t_{N_1}} 
=
\sum_{N_1\geq 0}\frac{q^{\, N_1^2}}{(q)_{N_1}},
\end{equation}

where $N_1$ again counts the number of all peaks and valleys. 

\subsubsection{Higher-$k$ models}

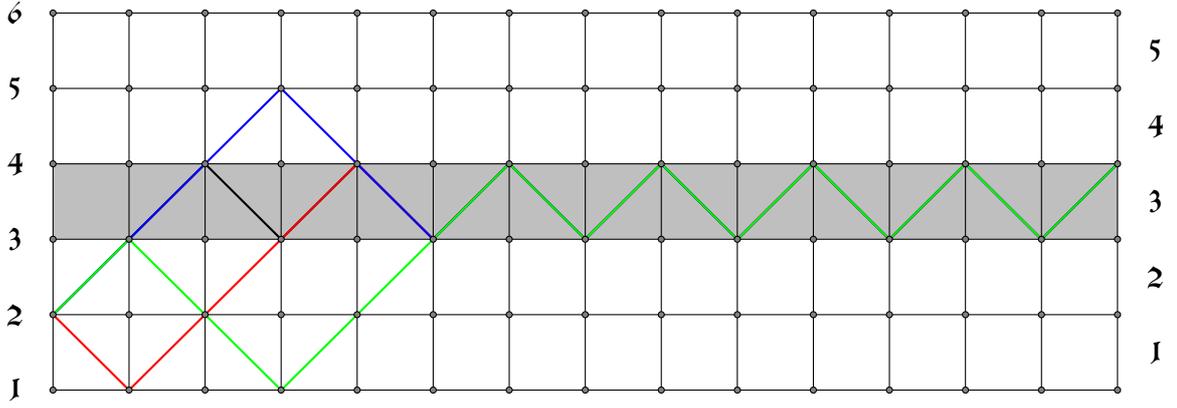
\begin{figure}
\begin{tikzpicture}[scale=1.0]

\draw [thin] (0.00, 5.0)--(14.00, 5.0);
\draw [thin] (0.00, 4.0)--(14.00, 4.0);
\draw [thin] (0.00, 3.0)--(14.00, 3.0);
\draw [thin] (0.00, 2.0)--(14.00, 2.0);
\draw [thin] (0.00, 1.0)--(14.00, 1.0);
\draw [thin] (0.00, 0.0)--(14.00, 0.0);

\draw [thin] ( 0.0, 0.0)--( 0.0, 5.0);
\draw [thin] ( 1.0, 0.0)--( 1.0, 5.0);
\draw [thin] ( 2.0, 0.0)--( 2.0, 5.0);
\draw [thin] ( 3.0, 0.0)--( 3.0, 5.0);
\draw [thin] ( 4.0, 0.0)--( 4.0, 5.0);
\draw [thin] ( 5.0, 0.0)--( 5.0, 5.0);
\draw [thin] ( 6.0, 0.0)--( 6.0, 5.0);
\draw [thin] ( 7.0, 0.0)--( 7.0, 5.0);
\draw [thin] ( 8.0, 0.0)--( 8.0, 5.0);
\draw [thin] ( 9.0, 0.0)--( 9.0, 5.0);
\draw [thin] (10.0, 0.0)--(10.0, 5.0);
\draw [thin] (11.0, 0.0)--(11.0, 5.0);
\draw [thin] (12.0, 0.0)--(12.0, 5.0);
\draw [thin] (13.0, 0.0)--(13.0, 5.0);
\draw [thin] (14.0, 0.0)--(14.0, 5.0);

\draw [thick] (0.0, 1.0)--(1.0, 2.0);
\draw [thick] (1.0, 2.0)--(2.0, 3.0);
\draw [thick] (2.0, 3.0)--(3.0, 2.0);
\draw [thick] (3.0, 2.0)--(4.0, 3.0);
\draw [thick] (4.0, 3.0)--(5.0, 2.0);
\draw [thick] (5.0, 2.0)--(6.0, 3.0);
\draw [thick] (6.0, 3.0)--(7.0, 2.0);
\draw [thick] (7.0, 2.0)--(8.0, 3.0);
\draw [thick] (8.0, 3.0)--(9.0, 2.0);
\draw [thick] (9.0, 2.0)--(10.0, 3.0);
\draw [thick] (10.0, 3.0)--(11.0, 2.0);
\draw [thick] (11.0, 2.0)--(12.0, 3.0);
\draw [thick] (12.0, 3.0)--(13.0, 2.0);
\draw [thick] (13.0, 2.0)--(14.0, 3.0);

\draw [thick,red] (0.0, 1.0)--(1.0, 0.0);
\draw [thick,red] (1.0, 0.0)--(2.0, 1.0);
\draw [thick,red] (2.0, 1.0)--(3.0, 2.0);
\draw [thick,red] (3.0, 2.0)--(4.0, 3.0);
\draw [thick,red] (4.0, 3.0)--(5.0, 2.0);
\draw [thick,red] (5.0, 2.0)--(6.0, 3.0);
\draw [thick,red] (6.0, 3.0)--(7.0, 2.0);
\draw [thick,red] (7.0, 2.0)--(8.0, 3.0);
\draw [thick,red] (8.0, 3.0)--(9.0, 2.0);
\draw [thick,red] (9.0, 2.0)--(10.0, 3.0);
\draw [thick,red] (10.0, 3.0)--(11.0, 2.0);
\draw [thick,red] (11.0, 2.0)--(12.0, 3.0);
\draw [thick,red] (12.0, 3.0)--(13.0, 2.0);
\draw [thick,red] (13.0, 2.0)--(14.0, 3.0);

\draw [thick,blue] (0.0, 1.0)--(1.0, 2.0);
\draw [thick,blue] (1.0, 2.0)--(2.0, 3.0);
\draw [thick,blue] (2.0, 3.0)--(3.0, 4.0);
\draw [thick,blue] (3.0, 4.0)--(4.0, 3.0);
\draw [thick,blue] (4.0, 3.0)--(5.0, 2.0);
\draw [thick,blue] (5.0, 2.0)--(6.0, 3.0);
\draw [thick,blue] (6.0, 3.0)--(7.0, 2.0);
\draw [thick,blue] (7.0, 2.0)--(8.0, 3.0);
\draw [thick,blue] (8.0, 3.0)--(9.0, 2.0);
\draw [thick,blue] (9.0, 2.0)--(10.0, 3.0);
\draw [thick,blue] (10.0, 3.0)--(11.0, 2.0);
\draw [thick,blue] (11.0, 2.0)--(12.0, 3.0);
\draw [thick,blue] (12.0, 3.0)--(13.0, 2.0);
\draw [thick,blue] (13.0, 2.0)--(14.0, 3.0);

\draw [thick,green] (0.0, 1.0)--(1.0, 2.0);
\draw [thick,green] (1.0, 2.0)--(2.0, 1.0);
\draw [thick,green] (2.0, 1.0)--(3.0, 0.0);
\draw [thick,green] (3.0, 0.0)--(4.0, 1.0);
\draw [thick,green] (4.0, 1.0)--(5.0, 2.0);
\draw [thick,green] (5.0, 2.0)--(6.0, 3.0);
\draw [thick,green] (6.0, 3.0)--(7.0, 2.0);
\draw [thick,green] (7.0, 2.0)--(8.0, 3.0);
\draw [thick,green] (8.0, 3.0)--(9.0, 2.0);
\draw [thick,green] (9.0, 2.0)--(10.0, 3.0);
\draw [thick,green] (10.0, 3.0)--(11.0, 2.0);
\draw [thick,green] (11.0, 2.0)--(12.0, 3.0);
\draw [thick,green] (12.0, 3.0)--(13.0, 2.0);
\draw [thick,green] (13.0, 2.0)--(14.0, 3.0);

\foreach \x in {0.0,...,14.0}
{
\draw [fill=black!50] (\x,5) circle (0.04);
\draw [fill=black!50] (\x,4) circle (0.04);
\draw [fill=black!50] (\x,3) circle (0.04);
\draw [fill=black!50] (\x,2) circle (0.04);
\draw [fill=black!50] (\x,1) circle (0.04);
\draw [fill=black!50] (\x,0) circle (0.04);
}

\node at (-0.50, 5.00) {$\textswab{6}$};
\node at (-0.50, 4.00) {$\textswab{5}$};
\node at (-0.50, 3.00) {$\textswab{4}$};
\node at (-0.50, 2.00) {$\textswab{3}$};
\node at (-0.50, 1.00) {$\textswab{2}$};
\node at (-0.50, 0.00) {$\textswab{1}$};

\node at (14.50, 4.50) {$\textswab{5}$};
\node at (14.50, 3.50) {$\textswab{4}$};
\node at (14.50, 2.50) {$\textswab{3}$};
\node at (14.50, 1.50) {$\textswab{2}$};
\node at (14.50, 0.50) {$\textswab{1}$};

\begin{scope}[on background layer]
\fill[gray!50] 
(0.00, 3.00) -- (14.00, 3.00) -- (14.00, 2.00) -- (0.00, 2.00);
\end{scope}

\end{tikzpicture}
\caption{\textit{
Several paths in the next-to-vacuum module of the restricted solid-on-solid 
model $\cL^{\, 2, 7}$, labeled by $a=2$, $b=3$, and $c=4$. 
The minimal path in black, the path with weight $1$ in red, and two paths 
with weight $3$ in blue and green. 
}}
\label{figure.path.02.next}
\end{figure}

A similar analysis extends to higher-$k$ models. We take $k=2$ as an example 
again. Here, we also have a possible new valley of weight 1 (see the red path 
in Figure \ref{figure.path.02.next}), and there are two types of particles 
(two paths with the same weight $wt=3$ are shown in Figure 
\ref{figure.path.02.next} in blue and green). These modifications are reflected 
in the constant-sign sum expression for the character 

\begin{equation}
    \sum_{N_1\geq N_2\geq0} 
    \frac{q^{\, N_1^2+N_2^2+N_2}}{(q)_{N_1-N_2}(q)_{N_2}}
    =
    1 + \frac{q}{1-q} + \frac{q^3}{1-q} + \cdots
\end{equation}

It still holds that $N_1$ counts the total number of particles, and $N_2$ counts 
the number of particles of the type shown in blue in Figure \ref{figure.path.02.next}. 

\subsection{A \texorpdfstring{$T$}{T}-refinement of the constant-sign sum expressions of 
the Virasoro characters as Macdonald indices}

Since there are $k$ particle species (as peaks or valleys) in the 
$\cL^{\, 2, \, 2k+3}$ model, a natural $t$-refined counting assigns a 
power of the refinement parameter $T$ to each particle, where $T = t / q$. 
In the Lee-Yang model $(k=1)$, there is only one type of particles, so 
we assign a weight $T$ to each particle in a path, and then the $t$-refined 
characters (written in terms of $T$) of the vacuum module and the next-to-vacuum module are   

\begin{equation}
\chi^{\, (2,5)}_{a=1}(q,T) =
\sum_{N_1\geq0} \frac{q^{\, N_1^2+N_1}}{(q)_{N_1}}T^{\, N_1},
\quad 
\chi^{\, (2,5)}_{a=2}(q,T)=\sum_{N_1\geq0} \frac{q^{\, N_1^2}}{(q)_{N_1}}T^{\, N_1}
\end{equation}

In the case of $k=3$, there are two types of (excited) particles with weight larger 
than $3$ in the vacuum module, and two in the next-to-vacuum module. We assign the 
weight $T$ to the first type that is counted by $N_1-N_2$ (such as the valley in 
Figure \ref{figure.path.07}, and the valley in green in 
Figure \ref{figure.path.02.next}), and $T^2$ to the other type counted by $N_2$ 
(such as the peak in Figure \ref{figure.path.06} and the peak in blue in 
Figure \ref{figure.path.02.next}). The $t$-refined character formulas are thus 
given by 

\begin{equation}
\chi^{\, (2,7)}_{a=1} \ll q, T \rr
=
\sum_{N_1\geq N_2\geq0} 
\frac{
q^{\, N_1^2 + N_2^2 + N_1 + N_2}
}{
(q)_{N_1-N_2}(q)_{N_2}} \, T^{\, N_1+N_2},
\quad 
\chi^{\, (2,7)}_{a=2}(q,T)
=
\sum_{N_1 \geq N_2 \geq 0} 
\frac{q^{\, N_1^2+N_2^2+N_2}}{(q)_{N_1-N_2}(q)_{N_2}}T^{\, N_1+N_2}
\end{equation}

We see in this way that in general we can refine the character with the factor 
\begin{equation}
    T^{\sum_{i=1}^{\, k}N_i},
\end{equation}

in the sum expression, that is to say, each particle of the $i$-th type, whose number 
is counted by $N_i-N_{i+1}$ (with $N_{k+1}\equiv 0$), is assigned a weight $T^i$. 
We remark that $\sum_{i=1}^{\, k}N_i$ is the linear part of the power of $q$, 
that is, $\sum_i N_i^2+N_i$, in the constant sign sum expression for the vacuum 
character. We will see from the series expansion of the sum expression that the above 
prescription matches Song's prescription to refine the Schur index to the Macdonald 
index, which also matches the computation of the Macdonald index from the TQFT approach. 

\section{A proposal for a closed-form expression for the Macdonald index}
\label{sec:proposal}
\textit{
We give our main proposals in the form of three conjectures and provide 
evidence for them. 
}

\subsection{Main proposal}

Recall that the $q$-series identities of Andrews--Gordon \cite{andrews.01,gordon}
take the form

\begin{equation}
\label{AG-2}
\chi^{\, (2,2k+3)}_{a}(q)=\sum_{N_1 \ge \cdots \ge N_{k} \ge 0}
\frac{ q^{\,  N_{1}^2 + \cdots + N_{k}^2 + N_{a} + \cdots + N_{k} }}
     {(q)_{N_{1}-N_{2}} \cdots (q)_{N_{k-1}-N_{k}} (q)_{N_{k}}}=\prod_{\begin{subarray}{c}
  n=1\\n\not\equiv0, \pm i \, (\mod 2k+3) \end{subarray}}^{\, \infty}
\frac{1}{1-q^n} ,
\end{equation}

where $|q|<1$, $k\ge1$ and $1\le a\le k+1$. We have already seen that $N_i$'s for 
$i=1, \cdots,k$ count the number of particles of different species in the paths 
approach. The $t$-refined version of the character (\ref{AG-2}), following the prescription 
we described in the previous section, then is 

\begin{equation}
\label{AG-3}
\chi^{\, (2,2k+3)}_{a}(q,T) =
\sum_{N_1 \ge \cdots \ge N_{k} \ge 0}
\frac{ 
q^{\,  N_{1}^2 + \cdots + N_{k}^2 + N_{a} + \cdots + N_{k} }
}{
(q)_{N_{1}-N_{2}} \cdots (q)_{N_{k-1}-N_{k}} (q)_{N_{k}}}
     T^{\, \sum_{i=1}^{\, k} N_i
}
\end{equation}

We first conjecture that the $t$-refined characters of the vacuum and next-to-vacuum 
modules are equal to the corresponding Macdonald indices for $n=2$. 

\subsubsection{Conjecture 1}

\begin{subequations}
\begin{equation}
    \chi^{\, (2,2k+3)}_{a=0}(q,T)=\cI_{(A_{1},A_{2k})}(q,t)=\sum_\lambda C^{\, -1}_\lambda(q,t)f^{\, I_{2,2k+1}}_\lambda(q,t),
\end{equation}
\begin{equation}
\chi^{\, (2,2k+3)}_{a=1}(q,T) = 
\cI^{\, \mathbb{S}^1}_{(A_{1},A_{2k})}(q,t) = 
\sum_\lambda C^{\, -1}_\lambda(q,t)f^{\, I_{2,2k+1}}_\lambda(q,t) \mathfrak{S}^1_\lambda(q,t)
\end{equation}
\end{subequations}

As there are series of fermionic sum expressions for characters of $\cW_3$ 
model with $(p, p')=(3, 7)$, we conjecture that the $t$-refined version of 
these expressions for the vacuum and next-to-vacuum modules agree with the 
corresponding Macdonald indices. 

\subsubsection{Conjecture 2} 

\begin{subequations}
\begin{multline}
    \chi^{\, (3,7)}_{(s_1,s_2)=(1,1)}(q,T)=\sum_{n_1, n_2, n_3, n_4 \geq 0}
\frac{
q^{\, 
(n_1 +   n_2 +   n_3)^2 + (n_2+n_3)^2 + {n_3}^2 + {n_4}^2 + (n_1 + 2 n_2 + 3 n_3) n_4 + 
(n_1 + 2 n_2 + 3 n_3 + 2 n_4)
}
}
{(q)_{n_1} (q)_{n_2} (q)_{n_3} (q)_{n_4}} 
\,
T^{\, n_1 + 2 n_2 + 3 n_3 + 2 n_4}\\
=\cI_{(A_{2},A_{3})}(q,t)=\sum_\lambda C^{\, -1}_\lambda(q,t)f^{\, I_{3,4}}_\lambda(q,t),
\end{multline}

\begin{multline}
    \chi^{\, (3,7)}_{(s_1,s_2)=(1,2)}(q,T)=\sum_{n_1, n_2, n_3, n_4 \geq 0}
\frac{
q^{\, 
(n_1 + n_2 + n_3)^2 + (n_2 + n_3)^2 + n_3^2 + n_4^2 + (n_1 + 2 n_2 + 3 n_3) n_4 + (n_2 + 2 n_3 + n_4)
}
}
{
(q)_{n_1} (q)_{n_2} (q)_{n_3} (q)_{n_4}
}
\, 
T^{\, n_1 + 2 n_2 + 3 n_3 + 2 n_4}\\
=\cI^{\, \mathbb{S}^{\, 1,0}}_{(A_{2},A_{3})}(q,t)=\sum_\lambda C^{\, -1}_\lambda(q,t)f^{\, I_{3,4}}_\lambda(q,t)\mathfrak{S}^{\, 1,0}_\lambda(q,t)
\end{multline}
\end{subequations}

We provide evidence for the above conjectures by comparing both sides as series expansions 
in $q$ (where $t = T q$) and matching them up to high orders. 

We further push this correspondence to interpret these particles as BPS operators contributing 
to the Schur/Macdonald index. 

\subsubsection{Conjecture 3: A path interpretation of aspects of the Schur index}

\begin{itemize}
\item The number of types of primary Schur operators is the number of particles
\footnote{\,
For definition of primary Schur operators, see \ref{primary.schur}
}.

\item Each path corresponds to a composite operator. 
Each particle in a path corresponds to a Schur operator. 

\item A particle at minimal position (smallest possible weight) 
corresponds to a primary Schur operator.
A particle far from a minimal position corresponds to a derivative 
of a primary Schur operator, that is, a descendant Schur operator 
\footnote{\,
For definition of descendant Schur operators, see \ref{primary.schur}
}, 
the distance from the minimal position equals the number of derivatives.
\end{itemize}

\subsection{The Macdonald version of the sum expressions of the Virasoro 
characters}

The expression for the $t$-refined Virasoro character is    
\begin{equation}
\label{AG-Mac}
\chi^{\, (2,2k+3)}_{a}(q,T)=\sum_{N_1 \ge \cdots \ge N_{k} \ge 0}
\frac{ 
q^{\,  N_{1}^2 + \cdots + N_{k}^2 + N_{a} + \cdots + N_{k} } 
}{
(q)_{N_{1}-N_{2}} \cdots (q)_{N_{k-1}-N_{k}} (q)_{N_{k}}
}
\, \, T^{\, N_{1} + \cdots + N_{k}}
\end{equation}

Let us list the $t$-refined characters of the vacuum module and the next-to-vacuum 
module for $k=1,2,3$ as a series expansion in $q$. 
\begin{multline}
\chi^{(2,5)}_{a=1}(q,T)=\sum_{N_1\geq0} \frac{q^{N_1^2+N_1}}{(q)_{N_1}}T^{N_1}
=1 + T q^2 + T q^3 + T q^4 + 
 T q^5 + \ll T + T^2 \rr \, q^6\\
 + \ll T + T^2 \rr \, q^7 + \ll T + 2 T^2 \rr \, q^8 + \ll T + 
    2 T^2 \rr \, q^9 + \ll T + 3 T^2 \rr \, q^{10}+\cO\ll \,q^{11}\, \rr ,\label{ref-ch-25-1}
\end{multline}

\begin{multline}
    \chi^{(2,5)}_{a=2}(q,T)=\sum_{N_1\geq0} \frac{q^{N_1^2}}{(q)_{N_1}}T^{N_1}=1 + T q + T q^2 + 
 T q^3 + \ll T + T^2 \rr \, q^4 + \ll T + T^2 \rr \, q^5 + \ll T + 2 T^2 \rr \, q^6\\
 + \ll T + 2 T^2 \rr \, q^7 + \ll T + 3 T^2 \rr \, q^8 + \ll T + 3 T^2 + T^3 \rr \, q^9 + \ll T + 
    4 T^2 + T^3 \rr \, q^{10}+\cO\ll \,q^{11}\, \rr ,\label{ref-ch-25-2}
\end{multline}

\begin{multline}
\chi^{(2,7)}_{a=1}(q,T)\\
=\sum_{N_1\geq N_2\geq0} \frac{q^{N_1^2+N_2^2+N_1+N_2}}{(q)_{N_1-N_2}(q)_{N_2}}T^{N_1+N_2}
=1 + T q^2 + 
 T q^3 + \ll T + T^2 \rr \, q^4 + \ll T + T^2 \rr \, q^5 + \ll T + 2 T^2 \rr \, q^6\\
 + \ll T + 2 T^2 \rr \, q^7 + \ll T + 3 T^2 + T^3 \rr \, q^8 + \ll T + 3 T^2 + 
    2 T^3 \rr \, q^9 + \ll T + 4 T^2 + 3 T^3 \rr \, q^{10}+\cO\ll \,q^{11}\, \rr ,\label{ref-ch-27-1}
\end{multline}

\begin{multline}
   \chi^{(2,7)}_{a=2}(q,T)\\
   = \sum_{N_1\geq N_2\geq0} \frac{q^{N_1^2+N_2^2+N_2}}{(q)_{N_1-N_2}(q)_{N_2}}T^{N_1+N_2}
    =1 + T q + 
 T q^2 + \ll T + T^2 \rr \, q^3 + \ll T + 2 T^2 \rr \, q^4 + \ll T + 2 T^2 \rr \, q^5\\
 + \ll T +  3 T^2 + T^3 \rr \, q^6
    + \ll T + 3 T^2 + 2 T^3 \rr \, q^7 + \ll T + 4 T^2 + 
    3 T^3 \rr \, q^8 \\
    + \ll T + 4 T^2 + 5 T^3 \rr \, q^9 + \ll T + 5 T^2 + 6 T^3 + 
    T^4 \rr \, q^{10}+\cO\ll \,q^{11}\, \rr ,\label{ref-ch-27-2}
\end{multline}

\begin{multline}
    \chi^{(2,9)}_{a=1}(q,T)\\
    =\sum_{N_1\geq N_2\geq N_3\geq0} \frac{q^{N_1^2+N_2^2+N_3^2+N_1+N_2+N_3}}{(q)_{N_1-N_2}(q)_{N_2-N_3}(q)_{N_3}}T^{N_1+N_2+N_3}
    =1 + T q^2 + 
 T q^3 + \ll T + T^2 \rr \, q^4 + \ll T + T^2 \rr \, q^5\\
 + \ll T + 2 T^2 + T^3 \rr \, q^6 
 + \ll T + 2 T^2 + T^3 \rr \, q^7 + \ll T + 3 T^2 + 2 T^3 \rr \, q^8 + \ll T + 3 T^2 + 
    3 T^3 \rr \, q^9 \\
    + \ll T + 4 T^2 + 4 T^3 + T^4 \rr \, q^{10}+\cO\ll \,q^{11}\, \rr ,\label{ref-ch-29-1}
\end{multline}

\begin{multline}
    \chi^{(2,9)}_{a=2}(q,T)\\
    =\sum_{N_1\geq N_2\geq N_3\geq0} \frac{q^{N_1^2+N_2^2+N_3^2+N_2+N_3}}{(q)_{N_1-N_2}(q)_{N_2-N_3}(q)_{N_3}}T^{N_1+N_2+N_3}=1 + T q + 
 T q^2 + \ll T + T^2 \rr \, q^3 + \ll T + 2 T^2 \rr \, q^4 \\
 + \ll T + 2 T^2 + 
    T^3 \rr \, q^5 + \ll T + 3 T^2 + 2 T^3 \rr \, q^6 
    + \ll T + 3 T^2 + 
    3 T^3 \rr \, q^7 
    + \ll T + 4 T^2 + 4 T^3 + T^4 \rr \, q^8 \\
    + \ll T + 4 T^2 + 6 T^3 + 
    2 T^4 \rr \, q^9 + \ll T + 5 T^2 + 7 T^3 + 4 T^4 \rr \, 
    q^{10}+\cO\ll \,q^{11}\, \rr \label{ref-ch-29-2}
\end{multline}
We will see later, while comparing with the Macdonald indices computed 
from the TQFT picture of gauge theories, that they agree with the above 
expressions (\ref{ref-ch-25-1})-(\ref{ref-ch-29-2}). 

\subsection{The sum expressions of the \texorpdfstring{$\cW_3$}{w3} characters}

The sum expressions for $(p,p')=(3,7)$ $\cW_3$ minimal models are developed in 
\cite{feigin.foda.welsh}, and take the form 

\begin{subequations}
\begin{multline}
\chi^{(3,7)}_{(s_1,s_2)=(1,1)}(q)=\sum_{n_1, n_2, n_3, n_4 \geq 0}
\frac{
q^{
(n_1 + n_2 + n_3)^2 + (n_2 + n_3)^2 + {n_3}^2 + {n_4}^2 + 
(n_1 + 2 n_2 + 3 n_3) n_4 + (n_1 + 2 n_2 + 3 n_3 + 2 n_4) 
}
}
{(q)_{n_1} (q)_{n_2} (q)_{n_3} (q)_{n_4}}\\
=\frac{1}{(q^2;q^7)(q^3;q^7)^2(q^4;q^7)^2(q^5;q^7)},\label{W3-11}
\end{multline}

\begin{multline}
\chi^{(3,7)}_{(s_1,s_2)=(1,3)}(q)=\sum_{n_1, n_2, n_3, n_4 \geq 0}
\frac{
q^{
(n_1 + n_2 + n_3)^2 + (n_2 + n_3)^2 + n_3^2 + n_4^2 + 
(n_1 + 2 n_2 + 3 n_3) n_4 + n_3 + n_4
}}
{
(q)_{n_1} (q)_{n_2} (q)_{n_3} (q)_{n_4}} \\
=\frac{1}{(q^1;q^7)(q^2;q^7)^2(q^5;q^7)^2(q^6;q^7)},\label{W3-13}
\end{multline}

\begin{multline}
\chi^{(3,7)}_{(s_1,s_2)=(2,2)}(q)=\sum_{n_1, n_2, n_3, n_4 \geq 0}
\frac{q^{
(n_1+n_2+n_3)^2 + (n_2+n_3)^2 + n_3^2 + n_4^2
+ (n_1 + 2 n_2 + 3 n_3) n_4
}}{
(q)_{n_1} (q)_{n_2} (q)_{n_3} (q)_{n_4}}\\
=\frac{1}{(q^1;q^7)^2(q^3;q^7)(q^4;q^7)(q^6;q^7)^2},\label{W3-22}
\end{multline}

\begin{multline}
\chi^{(3,7)}_{(s_1,s_2)=(1,2)}(q)=\sum_{n_1, n_2, n_3, n_4 \geq 0}
\frac{
q^{
(n_1+n_2+n_3)^2 + (n_2+n_3)^2 + n_3^2 + n_4^2 + (n_1 + 2 n_2 + 3 n_3) n_4 + (n_2 + 2 n_3 + n_4)     
}
}{
(q)_{n_1} (q)_{n_2} (q)_{n_3} (q)_{n_4}}\\
=\frac{1}{(q^1;q^7)(q^2(q^1;q^7);q^7)(q^3;q^7)(q^5;q^7)(q^1;q^7)(q^6;q^7)}\label{W3-12}
\end{multline}
\end{subequations}

The vacuum character (labeled by $s_1=s_2=1$) is given in equation (\ref{W3-11}) and 
the character of the next-to-vacuum module, labeled by $s_1=1$ and $s_2=2$, is given 
in (\ref{W3-12}). These two characters will be the main focus of ours in this model. 

We remark that as before, $n_{1,2,3,4}$ is the total number of particles 
of different species. This means that there are four types of fundamental particles 
in the $\cW_3$, $(p,p')=(3,7)$, model, and all states are composition of these 
fundamental particles or their (Schur) descendants following some selection rules. 
For example we can write the explicit form of the vacuum character, 
\begin{multline}
    \chi^{(3,7)}_{(s_1,s_2)=(1,1)}(q) =
    1+\frac{q^2}{1-q}+\frac{q^3}{1-q}+\frac{q^4}{1-q}+\frac{q^6}{1-q}+
    \frac{q^6}{(1-q)^2}+\frac{q^6}{(1-q)(1-q)^2}
    \\
    +\frac{q^8}{(1-q)(1-q)^2}+\frac{q^9}{(1-q)^2}+\frac{q^{10}}{(1-q)(1-q)^2}+ \cdots
\end{multline}
where the second term and the third term respectively correspond to $n_1=1$ and $n_4=1$ 
(other $n_i$'s being zero), and the sixth term $\frac{q^6}{(1-q)^2}$ is generated from 
$n_1=n_4=1$, $n_2=n_3=0$, that is the lowest contribution comes from the composition 
of a weight $2$ particle (counted by $n_1$) and a weight $4$ particle (counted by $n_4$). 

\subsection{The Macdonald version of the sum expressions of the \texorpdfstring{$\cW_3$}{w3} characters}

Now we consider the $T$-refinement of the fermionic characters (\ref{W3-11})-(\ref{W3-12}). As the path picture 
is currently not completely clear for higher-rank minimal models, the most natural 
generalization for $\cW_3$ is to add a refinement weight 

\begin{equation}
    T^{\, n_1 + 2 n_2 + 3 n_3 + 2 n_4},\label{ref-weight-W3}
\end{equation}

to each term in the summation, where $n_1 + 2 n_2 + 3 n_3 + 2 n_4$ is the linear term appearing 
in the power of $q$ in the vacuum character as in the case of $T$-refinement of Virasoro characters. 
In terms of particles, the refinement weight (\ref{ref-weight-W3}) means that we assign a weight $T$ to the 
first type counted by $n_1$, $T^2$ to the second type of particles counted by $n_2$ and so on. 
The refined expressions for each module in $(p,p')=(3,7)$ model are given below, 
together with their series expansions in $q$. 

\begin{subequations}
\begin{multline}
\label{W3-refine-1}
\chi^{(3,7)}_{(s_1,s_2)=(1,1)}(q,T)=\sum_{n_1, n_2, n_3, n_4 \geq 0}
\frac{
q^{
(n_1+n_2+n_3)^2 + (n_2+n_3)^2 + {n_3}^2 + {n_4}^2 + (n_1 + 2 n_2 + 3 n_3) n_4 + (n_1 + 2 n_2 + 3 n_3 + 2 n_4)
}
}
{(q)_{n_1} (q)_{n_2} (q)_{n_3} (q)_{n_4}} 
\, \, 
T^{\, n_1 + 2 n_2 + 3 n_3 + 2 n_4}\\
=1 + T q^2 + \ll T + T^2 \rr \, q^3 + \ll T + 2 T^2 \rr \, q^4 + \ll T + 2 T^2 \rr \, q^5 + \ll T + 
    3 T^2 + 2 T^3 \rr \, q^6 \\+ \ll T + 3 T^2 + 3 T^3 \rr \, q^7
    + \ll T + 4 T^2 + 5 T^3 + T^4 \rr \, q^8\\ + \ll T + 4 T^2 + 7 T^3 + 2 T^4 \rr \, q^9
    +\cO\ll \,q^{10}\, \rr 
\end{multline}

\begin{multline}
\label{W3-refine-4}
\chi^{(3,7)}_{(s_1,s_2)=(1,3)}(q,T)=\sum_{n_1, n_2, n_3, n_4 \geq 0}
\frac{
q^{
(n_1+n_2+n_3)^2 + (n_2+n_3)^2 + n_3^2 + n_4^2 + (n_1 + 2 n_2 + 3 n_3) n_4 + n_3 + n_4
}}
{
(q)_{n_1} (q)_{n_2} (q)_{n_3} (q)_{n_4}} 
T^{\, n_1 + 2 n_2 + 3 n_3 + 2 n_4}\\
= 1 + T q + \ll T + 2 T^2 \rr  q^2 + \ll T + 2 T^2 \rr  q^3 + \ll T + 3 T^2 + 
    2 T^3 \rr  q^4 + \ll T + 3 T^2 + 4 T^3 \rr  q^5 \\
    + \ll T + 4 T^2 + 6 T^3 + 
    2 T^4 \rr  q^6
    + \ll T + 4 T^2 + 8 T^3 + 4 T^4 \rr  q^7 \\+ \ll T + 5 T^2 + 
    10 T^3 + 9 T^4 \rr  q^8 
    + \ll T + 5 T^2 + 13 T^3 + 12 T^4 + 2 T^5 \rr  q^9
    +\cO\ll q^{10} \rr 
\end{multline}

\begin{multline}
\label{W3-refine-2}
\chi^{(3,7)}_{(s_1,s_2)=(2,2)}(q,T)=\sum_{n_1, n_2, n_3, n_4 \geq 0}
\frac{q^{
(n_1 + n_2 + n_3)^2 + (n_2 + n_3)^2 + n_3^2 + n_4^2 + (n_1 + 2 n_2 + 3 n_3) n_4
}}{
(q)_{n_1} (q)_{n_2} (q)_{n_3} (q)_{n_4}
}T^{\, n_1 + 2 n_2 + 3 n_3 + 2 n_4}\\
= 1 + \ll T + T^2 \rr  q + \ll T + 2 T^2 \rr  q^2 + \ll T + 2 T^2 + 2 T^3 \rr  q^3 + \ll T + 
    3 T^2 + 3 T^3 + T^4 \rr  q^4 \\
    + \ll T + 3 T^2 + 5 T^3 + 2 T^4 \rr  q^5 +
    \ll T + 4 T^2 + 7 T^3 + 5 T^4 \rr  q^6\\
    + \ll T + 4 T^2 + 9 T^3 + 8 T^4 + 
    2 T^5 \rr  q^7 + \ll T + 5 T^2 + 11 T^3 + 13 T^4 + 4 T^5 \rr  q^8 
    \\+ \ll T + 5 T^2 + 14 T^3 + 17 T^4 + 9 T^5 + T^6 \rr  q^9+\cO\ll q^{10} \rr 
\end{multline}

\begin{multline}
\label{W3-refine-3}
\chi^{(3,7)}_{(s_1,s_2)=(1,2)}(q,T)=\sum_{n_1, n_2, n_3, n_4 \geq 0}
\frac{
q^{
(n_1 + n_2 + n_3)^2 + (n_2 + n_3)^2 + n_3^2 + n_4^2 + (n_1 + 2 n_2 + 3 n_3) n_4 + (n_2 + 2 n_3 + n_4)
}
}
{
(q)_{n_1} (q)_{n_2} (q)_{n_3} (q)_{n_4}
}
\, 
T^{\, n_1 + 2 n_2 + 3 n_3 + 2 n_4}\\
= 1 + T q + \ll T + T^2 \rr \, q^2 + \ll T + 2 T^2 \rr \, q^3 + \ll T + 3 T^2 + 
    T^3 \rr \, q^4 + \ll T + 3 T^2 + 3 T^3 \rr \, q^5 \\
    + \ll T + 4 T^2 + 5 T^3 + 
    T^4 \rr \, q^6
    + \ll T + 4 T^2 + 7 T^3 + 2 T^4 \rr \, q^7 \\
    + \ll T + 5 T^2 + 9 T^3 + 
    6 T^4 \rr  q^8 + \ll T + 5 T^2 + 12 T^3 + 9 T^4 + T^5 \rr \, q^9+\cO\ll q^{10} \rr 
\end{multline}
\end{subequations}

\subsection{The ASW sum expressions of \texorpdfstring{$\cW_3$}{w3} characters} 

Another version of the sum expressions of the $\cW_3$, $(p,p')=(3,7)$, characters 
is proposed in \cite{andrews.schilling.warnaar},     

\begin{subequations}
\begin{equation}
\label{ASW-1}
\chi^{(3,7)}_{(s_1,s_2)=(1,1)}(q)=\sum_{n_1, n_2 \geq 0}
\frac{q^{n_1^2 - n_1 n_2 + n_2^2 + n_1 + n_2}}{(q)_{n_1}}
\qbinom{2n_1}{n_2} ,
\end{equation}

\begin{equation}
\label{ConjASW}
\chi^{(3,7)}_{(s_1,s_2)=(1,3)}(q)=\sum_{n_1, n_2 \geq 0}
\frac{q^{n_1^2 - n_1 n_2 +n_2^2 + n_2}}{(q)_{n_1}}
\qbinom{2 n_1 + 1}{n_2} ,
\end{equation}

\begin{equation}
\label{ASW-2}
\chi^{(3,7)}_{(s_1,s_2)=(2,2)}(q)=\sum_{n_1, n_2 \geq 0}
\frac{q^{n_1^2 -n_1 n_2 + n_2^2}}{(q)_{n_1}}
\qbinom{2n_1}{n_2} ,
\end{equation}

\begin{equation}
\label{ASW-3}
\chi^{(3,7)}_{(s_1,s_2)=(1,2)}(q)=\sum_{n_1,n_2\geq 0}
\frac{q^{n_1^2-n_1 n_2+n_2^2+n_1}}{(q)_{n_1}}\qbinom{2n_1+1}{n_2}
=
\sum_{n_1,n_2\geq 0}
\frac{q^{n_1^2-n_1 n_2+n_2^2+n_2}}{(q)_{n_1}}\qbinom{2n_1}{n_2},
\end{equation}
\end{subequations}
where 
\begin{equation}
    \qbinom{P}{N}=\lt \{\begin{array}{cc}
    \frac{(q)_P}{(q)_N(q)_{P-N}} & 0\leq N\leq P,\\
    0 & {\rm otherwise},\\
    \end{array} \rt.
\end{equation}
is the $q$-Gaussian polynomial
\footnote{\,
Equation (\ref{ConjASW}) has long time been a conjectured 
expression but was proved in \cite{corteel.welsh} recently.
}.

As before, we wish to refine the characters (\ref{ASW-1})-(\ref{ASW-3}) with $T$ to the power 
of the linear term in the power of $q$ in the numerator of the vacuum 
character, that is, $T^{n_1 + n_2}$. 
The Macdonald version of the $\cW_3$ characters that we obtain in this way 
are 

\begin{subequations}
\begin{multline}
\label{ASW-refine-1}
\chi^{(3,7)}_{(s_1,s_2)=(1,1)}(q,T)=\sum_{n_1, n_2 \geq 0}
\frac{q^{n_1^2 - n_1 n_2 + n_2^2 + n_1 + n_2}}{(q)_{n_1}}
\qbinom{2n_1}{n_2}T^{n_1+n_2}\\
=1 + T q^2 + \ll T + T^2 \rr \, q^3 + \ll T + 2 T^2 \rr \, q^4 + \ll T + 2 T^2 \rr \, q^5 + \ll T + 
    3 T^2 + 2 T^3 \rr \, q^6\\ + \ll T + 3 T^2 + 3 T^3 \rr \, q^7 
    + \ll T + 4 T^2 + 
    5 T^3 + T^4 \rr \, q^8 + \ll T + 4 T^2 + 7 T^3 + 2 T^4 \rr \, q^9+\cO\ll \,q^{10}\, \rr ,
\end{multline}

\begin{multline}
\label{ConjASW-1}
\chi^{(3,7)}_{(s_1,s_2)=(1,3)}(q,T)=\sum_{n_1, n_2 \geq 0}
\frac{q^{n_1^2 - n_1 n_2 +n_2^2 + n_2}}{(q)_{n_1}}
\qbinom{2 n_1 + 1}{n_2} T^{n_1+n_2}\\
=1 + T q + \ll 1 + 2 T \rr \, q^2 + 
 3 T q^3 + \ll 4 T + 2 T^2 \rr \, q^4 + \ll 5 T + 3 T^2 \rr \, q^5 + \ll 6 T + 
    7 T^2 \rr \, q^6\\ + \ll 7 T + 10 T^2 \rr \, q^7 
    + \ll 7 T + 17 T^2 + 
    T^3 \rr \, q^8 + \ll 7 T + 22 T^2 + 4 T^3 \rr \, q^9+\cO\ll \,q^{10}\, \rr  ,
\end{multline}

\begin{multline}
\label{ASW-refine-2}
\chi^{(3,7)}_{(s_1,s_2)=(2,2)}(q,T)=\sum_{n_1, n_2 \geq 0}
\frac{q^{n_1^2 -n_1 n_2 + n_2^2}}{(q)_{n_1}}
\qbinom{2n_1}{n_2} T^{n_1+n_2}\\
=1 + \ll T + T^2 \rr \, q + \ll T + 2 T^2 \rr \, q^2 + \ll T + 2 T^2 + 2 T^3 \rr \, q^3 + \ll T + 
    3 T^2 + 3 T^3 + T^4 \rr \, q^4 \\+ \ll T + 3 T^2 + 5 T^3 + 2 T^4 \rr \, q^5
    + \ll T + 
    4 T^2 + 7 T^3 + 5 T^4 \rr \, q^6\\
    + \ll T + 4 T^2 + 9 T^3 + 8 T^4 + 
    2 T^5 \rr \, q^7 + \ll T + 5 T^2 + 11 T^3 + 13 T^4 + 4 T^5 \rr \, q^8 \\
    + \ll T + 
    5 T^2 + 14 T^3 + 17 T^4 + 9 T^5 + T^6 \rr \, q^9+\cO(\,q^{10}\,) ,
\end{multline}

\begin{multline}
\label{ASW-refine-3}
\chi^{(3,7)}_{(s_1,s_2)=(1,2)}(q)=\sum_{n_1,n_2\geq 0}
\frac{q^{n_1^2-n_1 n_2+n_2^2+n_1}}{(q)_{n_1}}\qbinom{2n_1+1}{n_2}T^{n_1+n_2}
=
\sum_{n_1,n_2\geq 0}
\frac{q^{n_1^2-n_1 n_2+n_2^2+n_2}}{(q)_{n_1}}\qbinom{2n_1}{n_2}T^{n_1+n_2}\\
=1 + T q + \ll T + T^2 \rr \, q^2 + \ll T + 2 T^2 \rr \, q^3 + \ll T + 3 T^2 + 
    T^3 \rr \, q^4 + \ll T + 3 T^2 + 3 T^3 \rr \, q^5 \\
    + \ll T + 4 T^2 + 5 T^3 + 
    T^4 \rr \, q^6 
    + \ll T + 4 T^2 + 7 T^3 + 2 T^4 \rr \, q^7 + \ll T + 5 T^2 + 9 T^3 + 
    6 T^4 \rr \, q^8 \\
    + \ll T + 5 T^2 + 12 T^3 + 9 T^4 + T^5 \rr \, q^9+\cO\ll \,q^{10}\, \rr 
\end{multline}
\end{subequations}
We observe that (\ref{ASW-refine-1}), (\ref{ASW-refine-2}) and (\ref{ASW-refine-3}) respectively 
match (\ref{W3-refine-1}), (\ref{W3-refine-2}) and (\ref{W3-refine-3}) as series expansions, 
while (\ref{ConjASW-1}) does not match (\ref{W3-refine-4}). Since we only consider vacuum and next-to-vacuum characters, (\ref{W3-refine-1}) and (\ref{W3-refine-3}), in this article, this disagreement is not important to us at the moment. 

\subsection{Matching the Virasoro infinite-series of vacuum characters}

Let us list the Macdonald indices obtained in \cite{song.01,song.02,watanabe.zhu}. 

\begin{multline}
    \cI^{(A_1,A_{2})}=1+Tq^2+Tq^3+Tq^4+Tq^5+\ll T+T^2 \rr \,q^6+\ll T+T^2 \rr \,q^7+\ll T+2T^2 \rr \,q^8\\
    +\ll T+2T^2 \rr \,q^9+\ll T+3T^2 \rr \,q^{10}+\cO\ll \,q^{11}\, \rr ,\label{index-12}
\end{multline}

\begin{multline}
    \cI^{(A_1,A_{4})}=1+Tq^2+Tq^3+\ll T+T^2 \rr \,q^4+\ll T+T^2 \rr \,q^5+\ll T+2T^2 \rr \,q^6+\ll T+2T^2 \rr \,q^7\\
+\ll T+3T^2+T^3 \rr \,q^8+\ll T+3T^2+2T^3 \rr \,q^9+\ll T+4T^2+3T^3 \rr \,q^{10}+\cO\ll \,q^{11}\, \rr ,
\label{index-14}
\end{multline}

\begin{multline}
    \cI^{(A_1,A_{6})}=1+Tq^2+Tq^3+\ll T+T^2 \rr \,q^4+\ll T+T^2 \rr \,q^5+\ll T+2T^2+T^3 \rr \,q^6\\+\ll T+2T^2+T^3 \rr \,q^7
+\ll T+3T^2+2T^3 \rr \,q^8+\ll T+3T^2+3T^3 \rr \,q^9\\
+\ll T+4T^2+4T^3+T^4 \rr \,q^{10}+\cO\ll \,q^{11}\, \rr 
\label{index-16}
\end{multline}
The above results (\ref{index-12}), (\ref{index-14}) and (\ref{index-16}) match 
the $t$-refined characters obtained from our path approach 
(\ref{ref-ch-25-1}), (\ref{ref-ch-27-1}) and (\ref{ref-ch-29-1})

\subsection{Matching the Virasoro infinite-series of next-to-vacuum characters}

Following \cite{watanabe.zhu}, the Macdonald indices corresponding to the next-to-vacuum modules,
computed by inserting a surface defect with vortex number $s'=1$, are 

\begin{multline}
    \cI^{\mathbb{S}^1}_{(A_1,A_2)}(q,t)=1+Tq+Tq^2+Tq^3+\ll T+T^2 \rr \,q^4+\ll T+T^2 \rr \,q^5+\ll T+2T^2 \rr \,q^6\\+\ll T+2T^2 \rr \,q^7
+\ll T+3T^2 \rr \,q^8+\ll T+3T^2+T^3 \rr \,q^9+\ll T+4T^2+T^3 \rr \,q^{10}+\cO\ll \,q^{11}\, \rr ,\label{Mac-12-1}
\end{multline}
\begin{multline}
    \cI^{\mathbb{S}^1}_{(A_1,A_4)}(q,t)=1+Tq+Tq^2+\ll T+T^2 \rr \,q^3+\ll T+2T^2 \rr \,q^4+\ll T+2T^2 \rr \,q^5\\
    +\ll T+3T^2+T^3 \rr \,q^6
+\ll T+3T^2+2T^3 \rr \,q^7+\ll T+4T^2+3T^3 \rr \,q^8\\+\ll T+4T^2+5T^3 \rr \,q^9
+\ll T+5T^2+6T^3+T^4 \rr \,q^{10}+\cO\ll \,q^{11}\, \rr ,\label{Mac-14-1}
\end{multline}
\begin{multline}
    \cI^{\mathbb{S}^1}_{(A_1,A_6)}(q,t)=1+Tq+Tq^2+\ll T+T^2 \rr \, q^3+\ll T+2T^2 \rr \,q^4+\ll T+2T^2+T^3 \rr \,q^5\\
    +\ll T+3T^2+2T^3 \rr \,q^6
+\ll T+3T^2+3T^3 \rr \,q^7+\ll T+4T^2+4T^3+T^4 \rr \,q^8\\+\ll T+4T^2+6T^3+2T^3 \rr \,q^9
+\ll T+5T^2+7T^3+4T^4 \rr \,q^{10}+\cO\ll \,q^{11}\, \rr 
\label{Mac-16-1}
\end{multline}
and they match (\ref{ref-ch-25-2}), (\ref{ref-ch-27-2}) and (\ref{ref-ch-29-2}) computed from 
the path approach. 

\subsection{Matching the \texorpdfstring{$\cW_3$}{w3} vacuum and next-to-vacuum characters}

The Macdonald indices for rank-two Argyres-Douglas theories are also computed in \cite{watanabe.zhu} 
\textit{via} the TQFT approach, and the indices corresponding to the next-to-vacuum module are also 
conjectured based on the Higgsing approach. In this way, we obtained 

\begin{multline}
    \cI^{(A_2,A_3)}(q,t)=1+Tq^2+\ll T+T^2 \rr \,q^3+\ll T+2T^2 \rr \,q^4+\ll T+2T^2 \rr \,q^5+\ll T+3T^2+2T^3 \rr \,q^6\\
+\ll T+3T^2+3T^3 \rr \,q^7+\ll T+4T^2+5T^3+T^4 \rr \,q^8+\ll T+4T^2+7T^3+2T^4 \rr \,q^9+\cO\ll \,q^{10}\, \rr ,\label{Mac-23-00}
\end{multline}

\begin{multline}
    \cI^{\mathbb{S}^{1,0}}_{(A_2,A_3)}(q,t)=1+Tq+\ll T+T^2 \rr \,q^2+\ll T+2T^2 \rr \, q^3+\ll T+3T^2+T^3 \rr \, q^4+\ll T+3T^2+3T^3 \rr \,q^5\\
+\ll T+4T^2+5T^3+T^4 \rr \,q^6+\ll T+4T^2+7T^3+2T^4 \rr \,q^7+\cO\ll \,q^8\, \rr 
\label{Mac-23-10}
\end{multline}

Interestingly, (\ref{Mac-23-00}) and (\ref{Mac-23-10}) respectively match with 
(\ref{W3-refine-1}) and (\ref{W3-refine-3}) (or equivalently (\ref{ASW-refine-1}) 
and (\ref{ASW-refine-3})) up to the order computed for the Macdonald index. 

\subsubsection{Remark} 
\textit{
The above indices (\ref{Mac-23-00}) and (\ref{Mac-23-10}) are computed in the TQFT appraoch only with 
the wavefunction $f^{I_{3,4}}_{\emptyset}(q,t)$ and $f^{I_{3,4}}_{(2,1)}(q,t)$, and 
are truncated at the level that is not affected by the next non-trivial contributions 
from $f^{I_{3,4}}_{(3,0)}(q,t)$ and $f^{I_{3,4}}_{(3,3)}(q,t)$. 
}

\subsection{Relation with Schur operators}

Here, we focus on the cases corresponding to Virasoro minimal models, where the paths picture
is well-understood. For the Lee-Yang model $\cL^{\, 2, 5}$, the vacuum character is    

\begin{equation}
\sum_{N_1\geq 0}\frac{q^{N_1^2+N_1}}{(q)_{N_1}}=\sum_{N_1\geq 0}\sum_{\substack{t_1,t_2,  \cdots,t_{N_1}\\t_{i+1}-t_i\geq 2,t_1\geq 2}}q^{\sum_{i=1}^{N_1}t_i},
\end{equation}
and its $t$-refined version is   
\begin{equation}
    \sum_{N_1\geq 0}\frac{q^{N_1^2+N_1}}{(q)_{N_1}}T^{N_1}=\sum_{N_1=0}^\infty T^{N_1}\sum_{\substack{t_1,t_2,  \cdots,t_{N_1}\\t_{i+1}-t_i\geq 2\\t_1\geq 2}}q^{\sum_{i=1}^{N_1}t_i}
    \label{rewrite-Lee-Yang}
\end{equation}

Let $\cO$ denote the primary Schur operator that corresponds 
to the contribution $T q^2$ in the Macdonald index. Each particle with weight 
$t_i$ corresponds to $t_i-2$ derivatives
\footnote{\,
$\sigma^\mu_{\alpha\dot{\alpha}}$ or more explicitly
$(\sigma^\mu)_{\mu=0}^3=(\textbf{1},\sigma^1,\sigma^2,\sigma^3)$ 
is the a collection of Pauli matrices that can be used to convert 
the representation of the SO(4) Lorentz group to the spinors of 
SU(2)$\times$SU(2). 
$\sigma^\mu_{+\dot{+}}$ is the top component of this matrix, 
as a Schur operator always has to be the highest-weight state 
in the representation of Lorentz group \cite{beem.01}.
} 
acting on $\cO$, that is, the operator
$(\sigma^\mu_{+\dot{+}}\partial_\mu)^{t_i-2}\cO$. 
A general composite Schur operator made from $N_1$ such building blocks, of the 
form $:\prod_{i=1}^{N_1}(\sigma^\mu_{+\dot{+}}\partial_\mu)^{t_i-2}\cO:$, then
corresponds to a path with $N_1$ particles of weight $t_i$. It is natural 
in this context to conjecture that there is only one primary Schur operator, 
$\cO$, in the $(A_1,A_2)$ theory. Due to the fermionic nature of the particles, 
$:\cO \cO:$, for example, is not allowed in the spectrum. This corresponds to 
the superselection rule in the OPE of Schur operators. 

Similarly, in the $\Delta=-\frac{1}{5}$ module of the Lee-Yang model $\cL^{\, 2, 5}$, 
we prepare an operator $\cJ$ that corresponds to the contribution $Tq$ in the Macdonald 
index, then all peaks and valleys in the statistical mechanical model (with weight $t_i$) 
correspond to a Schur operator $(\sigma^\mu_{+\dot{+}}\partial_\mu)^{t_i-1}\cJ$. Each path 
with several peaks and valleys represents a composite Schur operator as a product $:\prod_{i}(\sigma^\mu_{+\dot{+}}\partial_\mu)^{t_i-1}\cJ:$

The case of $\cL^{\, 2, 7}$ model is more interesting. In the vacuum module, 
we have two types of particles when the weight is larger than or equal to $4$. 
At level 4, we have a descendant Schur operator 
$(\sigma^\mu_{+\dot{+}}\partial_\mu)^{2}\cO$, which contributes $Tq^4$ 
to the Macdonald index, and a primary Schur operator 
$\hat{\cC}_{1(\frac{1}{2},\frac{1}{2})}\sim :\cO\cO:$, 
which has Macdonald weight $T^2q^4$. 
The contribution from $\hat{\cC}_{2(1,1)}\sim :\cO\cO\cO:$ is missing in 
the Macdonald index, which agrees with the argument for the vanishing of 
the OPE coefficient 
$\lambda[\cO,\hat{\cC}_{1(\frac{1}{2},\frac{1}{2})},\hat{\cC}_{2(1,1}]$ in \cite{agarwal:2018}. 
This superselection rules is easily understood in the language of paths. 

More generally, the vanishing of the OPE coefficient 
$\lambda[\cO,\hat{\cC}_{k(\frac{k}{2},\frac{k}{2})},
\hat{\cC}_{(k+1)(\frac{k+1}{2},\frac{k+2}{2}}]$ 
matches with the fact that there are only $k$ types of particles in 
the statistical mechanical model of paths, and supports our conjecture 
regarding the correspondence between the Schur operators and the paths. 

In the case of $(p,p')=(3,7)$, there are four types of particles in 
the fermionic sums (\ref{W3-11}) to (\ref{W3-12}). From the discussion 
of \cite{agarwal:2018} to the effect that $\cW^2$ is not included in 
the spectrum, where $\cW=\cC_{1(0,0)}$, etc., 
it is consistent to identify the four primary Schur operators as 
$\cO=\hat{\cC}_{0(0,0)}$, $\hat{\cC}_{1(\frac{1}{2},\frac{1}{2})}\sim :\cO^2:$, 
$\hat{\cC}_{2(1,1)}\sim :\cO^3:$, and $\cW=\cC_{1(0,0)}$, whose refinement weights are respectively 
$T$, $T^2$, $T^3$ and $T^2$. In particular, the weight $T^2$ for $\cW$ agrees with the prescription 
given in \cite{watanabe.zhu}. The consistency with previous works on the gauge theory side also 
suggests that the formulation of (\ref{W3-11}) to (\ref{W3-12}) is essentially a free theory approach. 

\section{Comments}\label{sec:comment}

\subsection{Surface operators and characters}
Only the Macdonald indices computed in \cite{watanabe.zhu} that correspond to the vacuum 
module or the next-to-vacuum module (that is, 
in the Virasoro case, the $(r=1, s=1)$ and $(r=1, s=2)$ modules, and 
in the $\cW_3$  case, the 
$(r_1, r_2, s_1, s_2) = (1, 1, 1, 1)$ and 
$(r_1, r_2, s_1, s_2) = (1, 1, 1, 2)$ modules), 
are observed to directly take the form of a $t$-refined character. The Macdonald 
indices for more complicated modules, obtained using the same method, contain 
negative contributions. It is not clear whether only the Macdonald indices of 
the vacuum and the next-to-vacuum module have a physical meaning as $t$-refined 
characters in the dual chiral algebra. 

\subsection{Refining the bosonic version of a character}
In the case of Virasoro characters, it is possible to $t$-refine the bosonic version 
of a character using the Bailey lattice method of \cite{agarwal.andrews.bressoud}
\footnote{\,
We thank O Warnaar for bringing this to our attention.
}.
However, The Bailey refinement is a complicated one, as it involves not just 
the parameter $t$, but also the Bailey sequences ${\alpha_n}$ and ${\beta_n}$, 
$n=0, 1, \cdots$. 
The ${\beta}$ sequence can be trivialized ($\beta_n = \frac{1}{(q)_n}, n=0, 1, \cdots$) to obtain 
the refined fermionic version that we want (so we know that this is the correct 
$t$-refinement, but the bosonic version will now involve 
the ${\alpha_n}$ sequence and becomes quite complicated. 
For that reason, it seems to us that there is no advantage to $t$-refining the bosonic 
version in the case of Virasoro characters, 
since we know the $t$-refined fermionic versions, and we expect that the situation can 
get only (much) more complicated in the case of $\cW_3$ algebras where very little, 
and more general $W_N$ algebras where nothing is known about the fermionic versions 
of the characters or the Bailey lattice.

\subsection{The works of Bourdier, Drukker and Felix}
In \cite{bourdier.drukker.felix.01, bourdier.drukker.felix.02}, 
Bourdier, Drukker and Felix observed that the Schur index of certain theories 
can be written in terms of the partition function of a gas of fermions on 
a circle. 
It is not clear to us at this stage whether the latter fermions are related to 
ours. However, it is also entirely possible that the results of 
\cite{bourdier.drukker.felix.01, bourdier.drukker.felix.02} can be t-refined to 
obtain Macdonald indices. Further discussion of this is beyond the scope of this
work. 

\subsection{The works of Beem, Bonetti, Meneghelli, Peelaers and Rastelli}
Our work is definitely restricted to Song's approach to the Macdonald indices 
in $W_N$ models. In that approach, Song basically constructs the bosonic 
version of the character. Moreover, our work is restricted to those characters 
that we know the fermionic version thereof. It is entirely possible that the 
approach of the recent works
\cite{bonetti.meneghelli.rastelli, beem.maneghelli.rastelli, beem.maneghelli.peelaers.rastelli} 
is the right one to compute the Macdonald index in closed form in all generality.

\subsection{Paths, particles, instantons, BPS states and the Bethe/Gauge 
correspondence}
The paths are combinatorial objects that naturally belong to the 
representation theory of Virasoro irreducible highest weight modules 
\footnote{\,
The corresponding objects in the case of $W_N$ irreducible highest 
weight modules are Young tableaux that obey specific conditions
\cite{foda.dasmahapatra, dasmahapatra}.}. 
Following McCoy and collaborators
\cite{berkovich, berkovich.mccoy.01, berkovich.mccoy.schilling, kedem.klassen.mccoy.melzer} 
on the fermionic 
expressions of the Virasoro characters, the paths are 
interpreted in terms of (quasi-)particles and (quasi-)momenta
\cite{foda.quano.01, foda.quano.02, foda.lee.pugai.welsh, 
foda.welsh}. 
Subsequently, attempts were made to obtain the fermionic 
expressions of more elaborate objects, such as the correlation 
functions in statistical mechanics, or the conformal blocks in 
2D conformal field theories without success \cite{mccoy.unpublished}. 

After the discovery of Nekrasov's instanton 
partition function and the AGT correspondence, it became clear 
from \cite{bershtein.foda} that the fermionic expressions of 
the 2D conformal blocks in Virasoro minimal models are the 
Nekrasov instanton partition functions, and that 
the particles on the statistical mechanics/conformal 
field theory side are in correspondence with the instantons
on the gauge theory side. 

What we obtain in this work is a correspondence of a different 
type: a correspondence between the 
particles and the BPS states in Argyres-Douglas theories 
on the gauge side. It is natural to speculate that 
the Bethe/Gauge correspondence of Nekrasov and Shatashvili 
\cite{nekrasov.shatashvili.01, nekrasov.shatashvili.02} lies 
behind the results that we have obtained in this work.

\subsection{The thermodynamic Bethe Ansatz}
Connections between the combinatorics of the thermodynamic Bethe Ansatz and 
the combinatorics encoded in the paths were made clear in \cite{berkovich.mccoy.01}, 
and further in \cite{foda.welsh.02, warnaar.01, welsh}.
We anticipate that the methods of the thermodynamic Bethe Ansatz can be used 
to compute physical quantities in Argyres-Douglas theories.

\section*{Acknowledgements}
We thank 
Jean-Emile Bourgine, 
Matthew Buican, 
Dongmin Gang, 
Ian Grojnowski, 
Ralph Kaufmann,
Hee-Chol Kim, 
Kimyeong Lee,
Wolfger Peelaers, 
Leonardo Rastelli, 
Jaewon Song, 
S Ole Warnaar, 
Akimi Watanabe and 
Trevor Welsh 
for comments, correspondence and discussions. 
RZ wishes to thank APCTP and KIAS for hospitality, where this work was finalized.


\begin{thebibliography}{99}

\bibitem{agarwal:2016}
P Agarwal, K Maruyoshi and J Song, 
\textit{$ \mathcal{N} $ =1 Deformations and RG flows of $ \mathcal{N} $ =2 SCFTs, 
part II: non-principal deformations},
Journal of High Energy Physics \textbf{12} (2016) 103, 
\texttt{arXiv:1610.05311}

\bibitem{agarwal:2018}
P Agarwal, S Lee and J Song, 
\textit{Vanishing OPE Coefficients in 4d $N=2$ SCFTs},
Journal of High Energy Physics \textbf{06} (2019) 102, 
\texttt{arXiv:1812.04743} 

\bibitem{agarwal.andrews.bressoud}
A Agarwal, G E Andrews and D Bressoud,
\textit{The Bailey lattice}, 
The Journal of the Indian Mathematical Society, New Series \textbf{51} 57--73 

\bibitem{aharony.01}
O Aharony, J Marsano, S Minwalla, K Papadodimas, and M van Raamsdonk,
\textit{The Hagedorn - deconfinement phase transition in weakly coupled 
large N gauge theories},
Advances in Theoretical and Mathematical Physics \textbf{8} (2004) 603--696,
\texttt{arXiv:0310285 [hep-th]}

\bibitem{andrews.baxter.forrester}
G E Andrews, R J Baxter, and P J Forrester,
\textit{Eight-vertex SOS model and generalized Rogers-Ramanujan-type identities}, 
Journal of Statistical Physics \textbf{35.3} (1984) 193--266

\bibitem{andrews.01}
G E Andrews,
\textit{An analytic generalization of the Rogers-Ramanujan identities for odd moduli},
Proceedings of the National Academy of Sciences USA \textbf{ 71} (1974), 4082--4085

\bibitem{andrews.02}
G E Andrews,
\textit{Multiple series Rogers-Ramanujan type identities},
Pacific Journal of Mathematics \textbf{114} (1984), 267--283

\bibitem{andrews.schilling.warnaar}
G E Andrews, A Schilling and S O Warnaar,
\textit{An A$_2$ Bailey lemma and Rogers--Ramanujan-type identities},
Journal of the American Mathematical Society \textbf{ 12} (1999) 677--702,
\texttt{arXiv:math/9807125}

\bibitem{bailey}
W N Bailey,
\textit{Some identities in combinatory analysis},
Proceedings of the London Mathematical Society (2) \textbf{49} (1947), 
421--435


\bibitem{beem.01}
C Beem, M Lemos, P Liendo, W Peelaers, L Rastelli and B C van Rees, 
\textit{Infinite chiral symmetry in four dimensions}, 
Communications in Mathematical Physics,  
\textbf{336} (2015) 1359,  
\texttt{arXiv:1312.5344}

\bibitem{beem.02}
C Beem, W Peelaers, L Rastelli and B C van Rees, 
\textit{Chiral algebras of class S}, 
Journal of High Energy Physics 
\textbf{05} (2015) 020 
\texttt{arXiv:1408.6522}

\bibitem{beem.maneghelli.rastelli}
C Beem, C Meneghelli, and L Rastelli, 
\textit{Free Field Realizations from the Higgs Branch}, 
Journal of High Energy Physics 
\textbf{09} (2019) 058 
\texttt{arXiv:1903.07624}

\bibitem{beem.maneghelli.peelaers.rastelli}
C Beem, and C Meneghelli, and W Peelaers, and L Rastelli, 
\textit{VOAs and rank-two instanton SCFTs} (2019),  
\texttt{arXiv:1907.08629}

\bibitem{berkovich}
A Berkovich, 
\textit{Fermionic counting of RSOS states and Virasoro character 
formulas for the unitary minimal series $M \ll \nu, \nu + 1 \rr$: 
Exact results},
Nuclear Physics \textbf{B 431.1--2} (1994) 315--348,
\texttt{hep-th/9403073}

\bibitem{berkovich.02}
A Berkovich and B M McCoy,
\textit{},
Letters in Mathematical Physics \textbf{37.1} (1996) 49-66,
\texttt{arxiv:q-alg/9601012.pdf}

\bibitem{berkovich.mccoy.01}
A Berkovich and B M McCoy, 
\textit{Continued Fractions and Fermionic Representations for Characters 
of $M \ll p, p^{\, \prime} \rr$  minimal models}, 
Letters in Mathematical Physics \textbf{37} (1996) 49--66, 
\texttt{arXiv:hep-th/9412030} 

\bibitem{berkovich.mccoy.schilling}
A Berkovich, B M McCoy and A Schilling,
\textit{Rogers-Schur-Ramanujan type identities for the $M(p, p')$ minimal 
models of conformal field theory},
Communications in Mathematical Physics \textbf{191} (1998), 325--395,
\texttt{arXiv:q-alg/9607020} 


\bibitem{bershtein.foda}
M Bershtein and O Foda, 
\textit{AGT, Burge pairs and minimal models}, 
Journal of High Energy Physics, \textbf{06} (2014) 177, 
\texttt{arXiv:1404.7075}

\bibitem{bonetti.meneghelli.rastelli}
F Bonetti, C Meneghelli, and L Rastelli,
\textit{VOAs labelled by complex reflection groups and 4d SCFTs}
Journal of High Energy Physics, \textbf{05} (2019) 155, 
\texttt{arXiv:1810.03612}

\bibitem{bourdier.drukker.felix.01}
J Bourdier, N Drukker, and J Felix, 
\textit{The exact Schur index of $\mathcal{N}=4$ SYM}, 
Journal of High Energy Physics, \textbf{11} (2015) 210, 
\texttt{arXiv:1507.08659}

\bibitem{bourdier.drukker.felix.02}
J Bourdier, N Drukker, and J Felix, 
\textit{The $\mathcal{N}=2$ Schur index from free fermions}, 
Journal of High Energy Physics, \textbf{01} (2016) 167, 
\texttt{arXiv:1510.07041}

\bibitem{bressoud}
D Bressoud, 
\textit{Lattice paths and the Rogers-Ramanujan identities}, 
in Proceedings of the international Ramanujan centenary conference, 
Madras (1987), K Alladi, Editor, 
Lecture Notes in Mathematics \textbf{1395}, Springer (1989).

\bibitem{buican.nishinaka.01}
M Buican and T Nishinaka, 
\textit{On the superconformal index of Argyres-Douglas theories}, 
Journal of Physics A, \textbf{A49} (2016) 015401, 
\texttt{arXiv:1505.05884}

\bibitem{buican.nishinaka.02}
M Buican and T Nishinaka, 
\textit{Argyres-Douglas Theories, the Macdonald Index, and an RG 
Inequality}, 
Journal of High Energy Physics, \textbf{02} (2016) 159, 
\texttt{arXiv:1505.05884}

\bibitem{buican.nishinaka.03}
M Buican and T Nishinaka, 
\textit{On Irregular Singularity Wave Functions and Superconformal Indices}, 
Journal of High Energy Physics, \textbf{09} (2017) 066, 
\texttt{arXiv:1705.07173}

\bibitem{creutzig.01}
T Creutzig, 
\textit{W-algebras for Argyres-Douglas theories},
\texttt{arXiv:1506.00265}

\bibitem{creutzig.02}
T Creutzig, 
\textit{Logarithmic W-algebras and Argyres-Douglas theories at higher rank}, 
Journal of High Energy Physics \textbf{  11} (2018) 188,
\texttt{arXiv:1809.01725}

\bibitem{cordova.shao}
C Cordova and S-H Shao, 
\textit{Schur Indices, BPS Particles, and Argyres-Douglas Theories},
Journal of High Energy Physics \textbf{  01} (2016) 040,
\texttt{arXiv:1506.00265}

\bibitem{cordova.gaiotto.shao1}
C Cordova, D Gaiotto and S-H Shao, 
\textit{Infrared Computations of Defect Schur Indices},
Journal of High Energy Physics \textbf{  11} (2016) 106,
\texttt{arXiv:1606.08429}

\bibitem{cordova.gaiotto.shao2}
C Cordova, D Gaiotto and S-H Shao, 
\textit{Surface Defect Indices and 2d-4d BPS States},
Journal of High Energy Physics \textbf{  12} (2017) 078,
\texttt{arXiv:1703.02525}

\bibitem{cordova.gaiotto.shao3}
C Cordova, D Gaiotto and S-H Shao, 
\textit{Surface Defects and Chiral Algebras},
Journal of High Energy Physics \textbf{ 05} (2017) 140,
\texttt{arXiv:1704.01955}

\bibitem{corteel.welsh}
S Corteel and T Welsh,
\textit{The $A_2$ Rogers-Ramanujan identities revisited},
(2019),  
\texttt{arXiv:1905.08343}

\bibitem{dasmahapatra}
S Dasmahapatra, 
\textit{On the combinatorics of row and corner transfer 
matrices of the $A^{(1)}_{n-1}$ restricted face models}, 
International Journal of Modern Physics \textbf{A12} (1997) 
3551--3586, 
\texttt{arXiv:hep-th/9512095}

\bibitem{foda.dasmahapatra}
S Dasmahapatra and O Foda, 
\textit{Strings, paths, and standard tableaux}, 
International Journal of Modern Physics \textbf{A13} (1998) 501, 
\texttt{arXiv:q-alg/9601011}

\bibitem{dedushenko}
M Dedushenko, 
\textit{From VOAs to short star products in SCFT}, (2019),  
\texttt{arXiv:1911.05741}

\bibitem{dedushenko.fluder}
M Dedushenko and M Fluder, 
\textit{Chiral Algebra, Localization, Modularity, Surface defects, 
And All That}, (2019), 
\texttt{arXiv:1904.02704}

\bibitem{dedushenko.wang}
M Dedushenko and Y Wang, 
\textit{$4d/2d\rightarrow 3d/1d$: A song of protected operator algebras},
(2019), 
\texttt{arXiv:1912.01006}
        
\bibitem{feigin.foda.welsh}
B Feigin, O Foda and T A Welsh, 
\textit{Andrews-Gordon identities from combinations of Virasoro characters}, 
Ramanujan Journal \textbf{17} (2008) 33--52,
\texttt{arXiv:math-ph/0504014 }

\bibitem{feigin.fuchs}
B L Feigin and D B Fuchs,
\textit{Verma modules over the Virasoro algebra},
Functional Analysis and Applications \textbf{ 17} (1983), 
241--242

\bibitem{fluder.song}
M Fluder and J Song, 
\textit{Four-dimensional Lens Space Index from Two-dimensional Chiral Algebra},
Journal of High Energy Physics \textbf{07} (2018) 073,
\texttt{arXiv:1710.06029}

\bibitem{fluder.longhi}
M Fluder and P Longhi,
\textit{An infrared bootstrap of the Schur index with surface defects},
Journal of High Energy Physics \textbf{09} (2019) 062,
\texttt{arXiv:1905.02724}

\bibitem{foda.01}
O Foda,
\textit{Unpublished}, (2018)

\bibitem{foda.02}
O Foda, 
\textit{Off-critical local height probabilities on a plane and critical 
partition functions on a cylinder}, 
Nuclear Physics B \textbf{928} (2018) 279--326, 
\texttt{arXiv:1711.03337}

\bibitem{foda.lee.pugai.welsh}
O Foda, K S M Lee, Y Pugai and T Welsh, 
\textit{Path generating transforms}, 
Contemp. Math. \textbf{254} (2000) 157--186, 
\texttt{arXiv:math/9810043}

\bibitem{foda.quano.01}
O Foda and Y-H Quano, 
\textit{Polynomial identities of the Rogers--Ramanujan type}, 
International Journal of Modern Physics \textbf{  A 10} (1997), 2291--2315,
\texttt{arXiv:hep-th/9407191}

\bibitem{foda.quano.02}
O Foda and Y-H Quano,
\textit{Virasoro character identities from the Andrews-Bailey construction},
International Journal of Modern Physics \textbf{A 12} (1997), 1651--1675,
\texttt{arXiv:hep-th/9408086}

\bibitem{foda.warnaar}
O Foda and S O Warnaar, 
\textit{A bijection which implies Melzer's polynomial identities: the $\chi_{1,1}^{(p, p+ 1)}$ case}, 
Letters in Mathematical Physics \text{36} 2 (1996) 145--155, 
\texttt{arXiv:hep-th/9501088}

\bibitem{foda.welsh}
O Foda and T A Welsh, 
\textit{On the combinatorics of Forrester-Baxter models}, 
Physical Combinatorics Kyoto, Japan (1999) 49--103

\bibitem{foda.welsh.02}
O Foda and T Welsh,
\textit{Unpublished}, (2002)

\bibitem{forrester.baxter}
P J Forrester and R J Baxter, 
\textit{Further exact solutions of the eight-vertex SOS model and generalizations 
of the Rogers-Ramanujan identities}, 
Journal of Statistical Physics \textbf{38.3} (1985) 435--472

\bibitem{gadde:2011-1}
A Gadde, L Rastelli, S S Razamat, S Shlomo and W Yan, 
\textit{The 4d Superconformal Index from q-deformed 2d Yang-Mills},
Phys. Rev. Lett. \textbf{ 106} (2011) 241602,
\texttt{arXiv:1104.3850}

\bibitem{gadde:2011-2}
A Gadde, L Rastelli, S S Razamat, S Shlomo and W Yan, 
\textit{Gauge Theories and Macdonald Polynomials},
Communications in Mathematical Physics \textbf{ 319} (2013) 147--193,
\texttt{arXiv:1110.3740}

\bibitem{gaiotto.rastelli.razamat}
D Gaiotto, L Rastelli and S S Razamat, 
\textit{Bootstrapping the superconformal index with surface defects},
Journal of High Energy Physics \textbf{ 01} (2013) 022,
\texttt{arXiv:1207.3577}

\bibitem{gasper.rahman}
G Gasper and M Rahman,
\textit{Basic Hypergeometric Series},
Encyclopedia of Mathematics and its Applications \textbf{  35},
Cambridge University Press, 1990

\bibitem{gordon}
B Gordon,
{\em A combinatorial generalization of the Rogers-Ramanujan identities},
American Journal of Mathematics \textbf{83} (1961), 393--399

\bibitem{kaufmann}
R Kaufmann, 
\textit{Pathspace Decompositions for the Virasoro Algebra and its Verma Modules}, 
Int. J. Mod. Phys. \textbf{A 10} (1995), 943--962, 
\texttt{arXiv:hep-th/9405041}

\bibitem{kedem.klassen.mccoy.melzer}
R Kedem, T R Klassen, B M McCoy and E Melzer,
\textit{Fermionic sum representations for conformal field theory characters}, 
Physics Letters \textbf{B 307} (1993), 68--76,
\texttt{arXiv:hep-th/9301046}

\bibitem{kellendonk.recknagel}
J Kellendonk and A Recknagel,
\textit{Virasoro Representations on Fusion Graphs}, 
Physics Letters \textbf{B 298} (1993) 329--334,
\texttt{arXiv:hep-th/9210007}

\bibitem{kellendonk.rosgen.varnhagen}
J Kellendonk, M R\"osgen and R Varnhagen,
\textit{Path Spaces and W-Fusion in Minimal Models}, 
Int. J. Mod. Phys. \textbf{A 9} (1994) 1009--1024,
\texttt{arXiv:hep-th/9301086}

\bibitem{Kinney:2005ej}
J Kinney, J Maldacena, S Minwalla and S Raju,
\textit{An Index for 4 dimensional super conformal theories},
Communications in Mathematical Physics \textbf{275} (2007) 209--254,
\texttt{arXiv:0510251 [hep-th]}

\bibitem{maruyoshi.song}
K Maruyoshi and J Song, 
\textit{Enhancement of Supersymmetry \textit{via} Renormalization Group Flow and the Superconformal Index},
Physical Review Letter \textbf{118} (2017) 151602, 
\texttt{arXiv:1606.05632} 

\bibitem{maruyoshi.song.02}
K Maruyoshi and J Song, 
\textit{$\mathcal{N}=1 $ deformations and RG flows of $\mathcal{N}=2 $ SCFTs},
Journal of High Energy Physics \textbf{02} (2017) 075, 
\texttt{arXiv:1607.04281} 

\bibitem{mccoy.unpublished}
B M McCoy, 
\textit{Private communication}, (c. 2000)


\bibitem{nekrasov.shatashvili.01}
N A Nekrasov and S L Shatashvili, 
\textit{Supersymmetric vacua and Bethe ansatz}, 
Nucl. Phys. Proc. Suppl. \textbf{192--193} (2009) 91--112
\texttt{arXiv:0901.4744} 

\bibitem{nekrasov.shatashvili.02}
N A Nekrasov and S L Shatashvili, 
\textit{Quantum integrability and supersymmetric vacua}, 
Nucl. Phys. Proc. Suppl. \textbf{177} (2009) 105--119
\texttt{arXiv:0901.4748} 

\bibitem{nishinaka.sasa.zhu}
T Nishinanika, S Sasa and R-D Zhu,
\textit{On the Correspondence between Surface Operators in 
Argyres-Douglas Theories and Modules of Chiral Algebra}, 
Journal of High Energy Physics 
\textbf{03} (2019) 091 
\texttt{arXiv:1811.11772}

\bibitem{pan.peelars.01}
Y Pan and W Peelaers, 
\textit{Schur correlation functions on $S^3\times S^1$}, 
Journal of High Energy Physics
\textbf{07} (2019) 013 
\texttt{arXiv:1903.03623}

\bibitem{pan.peelars.02}
Y Pan and W Peelaers, 
\textit{Deformation quantizations from vertex operator algebras}, 
(2019),
\texttt{arXiv:1911.09631}

\bibitem{rastelli.private.communication}
L Rastelli,
\textit{Private communication}, 
(2018) 

\bibitem{rocha-caridi}
A Rocha-Caridi,
\textit{Vacuum vector representations of the Virasoro algebra},
in 
\textit{\lq\lq Vertex Operators in Mathematics and Physics\rq\rq}
J Lepowsky, S Mandelstam and I M Singer, Editors, Springer, 1985

\bibitem{rogers}
L J Rogers,
\textit{Second memoir on the expansion of certain infinite products},
Proceedings of the London Mathematical Society \textbf{  25} (1894) 318--343;
\textit{On two theorems of combinatory analysis and some allied identities},
Proceedings of the London Mathematical Society (2) \textbf{  16} (1917) 315--336

\bibitem{rogers-ramanujan}
L J Rogers and S Ramanujan,
\textit{Proof of certain identities in combinatory analysis},
Proceedings of the Cambridge Philosophical Society \textbf{  19} (1919) 211--216

\bibitem{slater}
L J Slater,
\textit{Further identities of the Rogers-Ramanujan type},
Proceedings of the London Mathematical Society (2) \textbf{  54} (1952) 147--167

\bibitem{song.01}
J Song,
\textit{Superconformal indices of generalized Argyres-Douglas 
theories from 2d TQFT}, 
Journal of High Energy Physics \textbf{02} (2016) 045,
\texttt{arXiv:1509.06730 [hep-th]}

\bibitem{song.02}
J Song, 
\textit{Macdonald index and chiral algebra},
Journal of High Energy Physics 
\textbf{08} (2017) 44,
\texttt{arXiv:1612.08956}

\bibitem{song.xie.yan}
J Song, D xie and W Yan,
\textit{Vertex operator algebras of Argyres-Douglas theories from M5-branes}, 
Journal of High Energy Physics 
\textbf{12} (2017) 123,
\texttt{arXiv:1706.01607}

\bibitem{xie.yan.01}
D Xie and W Yan, 
\textit{$W$ algebra, Cosets and VOAs for 4d 
$\mathcal{N} = 2$SCFT from M5 branes}, (2019), 
\texttt{arXiv:1902.02838}

\bibitem{xie.yan.02}
D Xie and W Yan, 
\textit{Schur sector of Argyres-Douglas theory and $W$-algebra}, (2019),  
\texttt{arXiv:1904.09094}

\bibitem{xie.yan.03}
D Xie and W Yan, 
\textit{4d $\mathcal{N}=2$ SCFTs and lisse W-algebras}, (2019), 
\texttt{arXiv:1910.02281}

\bibitem{warnaar.01}
S O Warnaar, 
\textit{Fermionic solution of the Andrews-Baxter-Forrester model. 
I. Unification of TBA and CTM methods}, 
Journal of statistical physics, \textbf{82} 3--4 (1996) 657--685, 
\texttt{arXiv:hep-th/9501134}

\bibitem{warnaar.02}
S O Warnaar, 
\textit{Fermionic solution of the Andrews-Baxter-Forrester model. II. Proof of Melzer's polynomial identities}, 
Journal of statistical physics, \textbf{84} 1--2 (1996) 49--83, 
\texttt{arXiv:hep-th/9508079}

\bibitem{warnaar.03}
S O Warnaar,
\textit{The Andrews-Gordon identities and $q$-multinomial coefficients},
Communications in Mathematical Physics \textbf{184.1} (1997) 203-232,
\texttt{arxiv:q-alg/9601012}

\bibitem{warnaar}
S O Warnaar,
\textit{Hall--Littlewood functions and the A$_2$ Rogers-Ramanujan 
identities}, 
Advances in Mathematics \textbf{200} (2006) 403--434, 
\texttt{arXiv:math/0410592 [math.CO]}

\bibitem{watanabe.zhu}
A Watanabe and R-D Zhu,
\textit{Testing Macdonald Index as a Refined Character of Chiral 
Algebra}, (2019),  
\texttt{arXiv:1909.04074}

\bibitem{welsh}
T A Welsh,
\textit{Fermionic expressions for minimal model Virasoro characters},
Memoirs of the American Mathematical Society \textbf{  175} (no. 827) (2005), 
\texttt{math.CO/0212154}

\end{thebibliography}
\end{document}